\documentclass[10pt,journal,compsoc]{IEEEtran}

\usepackage{enumerate}
\usepackage{textcomp}
\usepackage{float}
\usepackage{booktabs}
\usepackage{multirow}

\usepackage{hyperref}
\usepackage[ruled,vlined]{algorithm2e}

\usepackage{amsmath,amssymb,amsfonts}
\ifCLASSOPTIONcompsoc
  \usepackage[nocompress]{cite}
\else
  \usepackage{cite}
\fi
\ifCLASSINFOpdf
   \usepackage[pdftex]{graphicx}
\else
\fi
\usepackage{xspace}

\newcommand{\parsec}{\textsc{PaRSEC}\xspace}

\newcommand{\starpu}{\textsc{StarPU}\xspace}

\newcommand{\nlopt}{\textsc{NLopt}\xspace}

\newcommand{\chameleon}{\textsc{Chameleon}\xspace}

\newcommand{\hicma}{\textsc{HiCMA}\xspace}
\newcommand{\starsh}{\textsc{STARS-H}\xspace}

\usepackage{algorithmic}

\usepackage{subcaption}

\begin{document}
%
\title{High Performance Multivariate Geospatial Statistics on Manycore Systems}
%
%
%
%

\author{Mary Lai O. Salva\~na, Sameh Abdulah, Huang Huang, Hatem Ltaief,\\ Ying Sun, Marc G. Genton, and David E. Keyes
\IEEEcompsocitemizethanks{\IEEEcompsocthanksitem The authors are with the Extreme Computing Research Center, Computer,
Electrical, and Mathematical Sciences and Engineering Division (CEMSE),
King Abdullah University of Science and Technology (KAUST), Thuwal
23955-6900, Saudi Arabia. E-mail: \{Marylai.Salvana, Sameh.Abdulah, Huang.Huang, Hatem.Ltaief, Ying.Sun, Marc.Genton, David.Keyes\}@kaust.edu.sa. \protect\\}}


%
%

\markboth{IEEE  TRANSACTIONS ON PARALLEL AND DISTRIBUTED SYSTEMS,~Vol.~XX, No.~XX, MM, YYYY}%
{Shell \MakeLowercase{\textit{et al.}}: Bare Advanced Demo of IEEEtran.cls for IEEE Computer Society Journals}
%



\IEEEtitleabstractindextext{%
\begin{abstract}
Modeling and inferring spatial relationships and predicting missing values of environmental 
data are some of the main tasks of geospatial statisticians. These routine tasks are accomplished 
using multivariate geospatial models and the cokriging technique. The latter requires the evaluation 
of the expensive Gaussian log-likelihood function, 
which has impeded the adoption of multivariate geospatial models 
for large multivariate spatial datasets.
However, this large-scale cokriging challenge provides a fertile 
ground for supercomputing implementations for the geospatial statistics community as it 
is paramount to scale computational capability to match the growth in environmental data 
coming from the widespread use of different data collection technologies. In this paper, 
we develop and deploy large-scale multivariate spatial modeling and inference on parallel 
hardware architectures. To tackle the increasing complexity in matrix operations
and the massive concurrency in parallel systems, we 
leverage low-rank matrix approximation techniques with task-based programming models
and schedule the asynchronous computational tasks using a dynamic runtime system. 
The proposed framework provides both the dense and the approximated computations of the Gaussian log-likelihood 
function. It demonstrates accuracy robustness and performance scalability on a variety of 
computer systems. Using both synthetic and real datasets, the low-rank matrix 
approximation shows better performance compared to exact computation, while preserving 
the application requirements in both parameter estimation and prediction accuracy. We 
also propose a novel algorithm to assess the prediction accuracy after the online parameter 
estimation. The algorithm quantifies prediction performance and provides a benchmark for 
measuring the efficiency and accuracy of several approximation techniques in multivariate 
spatial modeling.
\end{abstract}

\begin{IEEEkeywords}
Gaussian log-likelihood, geospatial statistics, high-performance computing, large multivariate spatial data, low-rank  approximation, multivariate modeling/prediction. 
\end{IEEEkeywords}}

\maketitle

\IEEEdisplaynontitleabstractindextext

%
\IEEEpeerreviewmaketitle

\ifCLASSOPTIONcompsoc
\IEEEraisesectionheading{\section{Introduction}\label{sec:introduction}}
\else
\section{Introduction}
\label{sec:introduction}
\fi

\IEEEPARstart{T}{he} convergence of high-performance computing (HPC) and big data brings
 great promise in accelerating and improving large-scale  applications~\cite{asch2018big, liao2018moving}
on climate and weather modeling~\cite{abdulah2018exageostat}, astronomy~\cite{doucet2019mixed}, transportation~\cite{park2018high}, and bioinformatics~\cite{chu2019high}. Climate and weather modeling,
in particular, is one of the first applications of HPC for big data \cite{drake1995introduction}. 
The need to improve climate and weather models has pushed for advances in environmental data
collection technologies such as spaceborne, airborne, and ground sensors \cite{national2000research}. 
The volume of data coming from these sources is huge and increasing. For instance, NASA's Earth
Observing System Data and Information System (EOSDIS) is expected to archive more than 37 
petabytes of data in 2020 \cite{esds}. By 2022, the yearly increase is projected at 47.7 petabytes. 

Environmental data, such as climate and weather variables, are often recorded from different
spatial locations, and thus indexed by $\mathbf{s} \in \mathbb{R}^d, d\geq 1$, where $\mathbf{s}$ is
the location of the measurement. Usually, there are multiple variables measured at each location, 
such as temperature, humidity, wind speed, and atmospheric pressure. These colocated variables
may or may not depend on each other and on the variables at other locations. 

Exposing spatial relationships among spatially referenced data can be accomplished 
using geographical information systems (GIS)\cite{burrough2001gis}.
Through GIS, one can produce scientific visualizations such as maps of raw data and spatial patterns, 
thereby facilitating exploratory data analysis, statistical analysis, and hypothesis testing. 
The big data era brought new challenges to GIS and its ability to process and analyze
huge streams and volumes of geospatial data. However, the influx of big geospatial data were 
met with brand new capabilities of GIS made possible by HPC \cite{shi2013modern}. 
HPC boosts GIS operations and computations in the face of large amounts of geospatial 
data by utilizing modern hardware architectures such as computer clusters, GPUs, and cloud
computing infrastructures \cite{liu2013higis, stojanovic2013high, zhang2013parallel, yang2020big}. 
For instance, in \cite{li2013utilizing}, GPU accelerators have been used to accelerate the visualization of large-scale geospatial data. In \cite{stojanovic2013high}, distributed GPU systems through Message Passing Interface (MPI) 
over Network of Workstations (NoW) and Compute Unified Device Architecture (CUDA) have been used to perform real-time map matching and slope computations of a large global positioning system (GPS) data. Another example has been shown in \cite{kang2018using} where a graph-based methodology on a cluster computing paradigm was employed in land use/land cover change (LUCC) analysis; see \cite{zhu2020next} for other HPC-based GIS implementations review.

Aside from processing a huge amount of data harvested from different sources such as 
satellite images, model simulations, sensors, and the Internet of Things,
a major concern when dealing with environmental datasets is missing data on one or a few variables.
For instance, when using environmental variables as inputs to climate and weather models, the gaps
in areas with no measurements caused by poor atmospheric conditions or defective sensors, to name
a few, need to be filled \cite{bancheri2018design, zhang2018missing}. Several missing data interpolation 
techniques exist in the literature including numerical models, machine learning and deep learning models, and
geospatial statistics models \cite{fouilloy2018solar, bergen2019machine, zhang2019joint, demyanov2020special, irrgang2021will}. 
Numerical models solve a complex set of partial differential
equations and generate large volumes of predictions on the quantities of interest, 
such as the concentrations of pollutants in the atmosphere \cite{binkowski2003models,minet2020quantifying}.
The strength in using numerical models in prediction lies in their physically consistent 
representations of the phenomena or systems being analyzed. However, numerical models 
cannot accurately predict on very fine spatial resolutions \cite{yu2020data}.
Machine learning and deep learning models capture the spatial relationships among
environmental variables through feature representation learning using neural networks \cite{amato2020novel}. Image inpainting with 
transfer learning has been used to impute missing global surface temperature 
measurements in HadCRUT4 \cite{kadow2020artificial}. Generative adversarial network (GAN) was used in 
\cite{kang2021restoration} to reconstruct the missing sea surface temperature 
readings in satellite images due to cloud disturbances. Several other machine learning and deep learning
solutions to the missing data problem are listed in \cite{yu2020data}. Although machine learning and deep learning 
have shown high prediction capabilities, they are devoid of physical knowledge of the 
process being modeled and predicted \cite{irrgang2021will}. Furthermore, they suffer the  drawback of being unable 
to describe explicitly the spatial relationship among environmental variables\cite{mckinley2020special}, i.e., absence of interpretability.

In this work, we adopt the multivariate geospatial statistical approach due to its ability to characterize the dependence structure of the underlying spatial data. The geospatial statistical models are also very easy to interpret and yield very high prediction accuracy.
Multivariate geospatial statistics can interpolate environmental variables at unsampled locations 
by modeling the multivariate spatial dependencies using a multivariate covariance function,
whose parameters are calibrated with the aid of a log-likelihood function.
While every variable of interest can be modeled and predicted separately,
it has been shown that more accurate predictions can be produced when modeling dependent 
variables jointly~\cite{genton2015cross, zhang2015doesn}.

Modeling the variables as realizations from a multivariate Gaussian random field is the cornerstone of
multivariate geospatial statistics. Multivariate random fields are the equivalent of multivariate random
variables, where a vector of variables is measured at each location \cite{hristopulos2020random}.
Mathematically, this means that at location $\mathbf{s} \in \mathbb{R}^d, \; d\geq 1$, each variable
is considered as one component of the $p$-dimensional vector $\mathbf{Z}(\mathbf{s})$, i.e., $\mathbf{Z}(\mathbf{s})=\left\{ Z_1(\mathbf{s}), \ldots, Z_p(\mathbf{s})\right\}^{\top}$, where $\top$ is the transpose operator, $p$ is
the number of variables, and $Z_i(\mathbf{s}) \in \mathbb{R}$ indicates the value of the $i$th 
variable at location $\mathbf{s}$, $i = 1, \ldots, p$. When $\mathbf{Z}(\mathbf{s})$ is Gaussian, 
it is completely characterized by its mean vector and multivariate covariance function, which is 
more commonly known as cross-covariance function in geospatial statistics and multivariate kernel
in computer science.

Cokriging is the prediction of multiple variables using an optimal predictor \cite{genton2015cross},\cite{zhang2015doesn}.
To perform cokriging on a multivariate Gaussian random field, one only needs to specify a mean vector
and a cross-covariance function. In this work, we assume that the multivariate Gaussian random field has mean
zero and focus our attention on the cross-covariance function, which introduces considerable computational
challenges. This zero-mean assumption is reasonable since regression can be used to remove the mean.
Furthermore, we restrict our attention to a specific cross-covariance function that is stationary and isotropic.
These assumptions on the cross-covariance function are building blocks to more sophisticated ones including nonstationary and anisotropic cross-covariance functions, which can be readily accommodated by our proposed
framework. They do not present significant limitations on the large-scale multivariate geospatial analysis aimed at this work.

In practice, a class of cross-covariance functions is first selected and its unknown parameters are estimated
from the data \cite{cressie2016multivariate, genton2015cross, li2011approach}. Estimation relies on the
maximum likelihood estimation (MLE) approach~\cite{mardia1984maximum,kaufman2008covariance}.
Suppose $\boldsymbol{\theta} \in \mathbb{R}^q$ collects all the $q$ true unknown parameters of the
cross-covariance function. The MLE of $\boldsymbol{\theta}$, denoted by $\widehat{\boldsymbol{\theta}} \in \mathbb{R}^q$, is the $q$-dimensional vector which maximizes the log-likelihood function
\begin{equation} \label{eqn:loglikelihood_function}
l(\boldsymbol{\theta}) = -\frac{n p}{2} \log (2 \pi) - \frac{1}{2} \log |\boldsymbol{\Sigma}(\boldsymbol{\theta})| - \frac{1}{2} \mathbf{Z}^{\top} \boldsymbol{\Sigma}(\boldsymbol{\theta})^{-1} \mathbf{Z},
\end{equation}
with respect to all $q$ parameters in $\boldsymbol{\theta}$. Here $\mathbf{Z} \in \mathbb{R}^{pn}$ 
collects all the $p$-dimensional vectors $\mathbf{Z}(\mathbf{s})$ at $n$ locations, $\mathbf{s}_1, \mathbf{s}_2, \ldots, \mathbf{s}_n$, i.e., $\mathbf{Z} = \left\{ \mathbf{Z}(\mathbf{s}_1)^{\top},  \mathbf{Z}(\mathbf{s}_2)^{\top}, \ldots, \mathbf{Z}(\mathbf{s}_n)^{\top} \right\}^{\top}$. $\boldsymbol{\Sigma}(\boldsymbol{\theta})$ is the $pn \times pn$ 
cross-covariance matrix for $\mathbf{Z}$ and $|\boldsymbol{\Sigma}(\boldsymbol{\theta})|$ denotes the 
determinant of $\boldsymbol{\Sigma}(\boldsymbol{\theta})$. The entries of $\boldsymbol{\Sigma}(\boldsymbol{\theta})$ 
are calculated from the cross-covariance function that is known up to $\boldsymbol{\theta}$. 
The procedure in constructing this cross-covariance matrix is discussed in more detail in Section~\ref{sec:exact_multivariate_modeling}.

The MLE involves the computation of the log-likelihood function in Equation (\ref{eqn:loglikelihood_function}) for
each iteration in the optimization. In large-scale multivariate problems with irregularly spaced locations, 
the log-likelihood requires $\mathcal{O}(p^2n^2)$ memory and $\mathcal{O}(p^3n^3)$ operations per 
iteration, due to the Cholesky decomposition of $\boldsymbol{\Sigma}(\boldsymbol{\theta})$.
The prohibitive computational cost and corresponding storage of computing the log-likelihood function can be
reduced by relying on low-rank approximation techniques that exploit the low-rank representations of the
cross-covariance matrix, for instance, the Tile Low-Rank (TLR) 
approximation \cite{akbudak2017tile, keyes2020hierarchical, abdulah2018parallel}.
TLR  is the preferred approximation approach in case of parallel execution and it involves dividing the matrix
 into a set of tiles, then applying low-rank approximation separately to each tile.
 
The quality of the predictions can be assessed using the mean square prediction error (MSPE)~\cite{abdulah2018parallel} and the mean square relative prediction error (MSRP)~\cite{yan2018gaussian}. However, it has been shown that these commonly used criteria cannot adequately assess the prediction efficiency when different approximation methods are involved and are up for comparisons~\cite{hong2019efficiency}. As an alternative, under univariate modeling, the authors in~\cite{hong2019efficiency} proposed two criteria, namely, the mean loss of efficiency (MLOE) and the mean misspecification of the mean square error (MMOM), to more appropriately assess the loss of prediction efficiency when an approximated version of the model was used instead of the exact one. In this work, we first present a parallel implementation based on our software stack of the univariate MLOE/MMOM criteria, and then we propose a modified algorithm that extends these criteria to assess the multivariate approximations modeling on large-scale multivariate spatial datasets.

The remainder of the paper is as follows. Section~\ref{sec:contributions} summarizes the contribution of this work. 
Section~\ref{sec:related_work} recalls some of the established approaches in 
large-scale multivariate geospatial modeling and inference.
Section~\ref{sec:overview_of_the_problem} contains a brief 
discussion on multivariate geospatial statistics, 
an introduction to the log-likelihood estimation problem, 
and a review to some approximation techniques 
 which ameliorate the complexities encountered in MLE operations. 
  Section~\ref{sec:proposed_framework} establishes the research contributions of this paper. 
  Section~\ref{sec:performance} provides detailed illustrations of our proposed modeling framework 
  on synthetic and real datasets. Section~\ref{sec:conclusion} delivers
   an overall summary and conclusion.

\section{Contributions}\label{sec:contributions}

The six-fold contributions of the paper are as follows:

\begin{itemize}
\item We present multivariate geospatial modeling and inference in large-scale systems on both exact and TLR-based approximation computations with reduced complexity on the log-likelihood for both estimation and spatial prediction.

\item We propose a novel multivariate assessment algorithm  based on existing univariate criteria to evaluate our TLR-based parameter estimation accuracy.

\item We implement a parallel version of the univariate and the new multivariate criteria to assess the prediction efficiency on synthetic and real datasets.

\item We port the proposed implementation on shared-memory, GPUs, and distributed-memory systems using a modular software stack.

\item We evaluate the performance of both the parallel exact and TLR-based MLE computations, as well as the proposed multivariate assessment criteria, using different parallel platforms such as Intel Xeon Skylake/Cascade Lake, AMD EPYC (Rome), ARM ThunderX2, NVIDIA V100 GPU, and a distributed-memory Cray XC40 system.

\item We conduct a set of  experiments designed to assess the accuracy of our implementation in terms of inference and prediction on both synthetic and real datasets.

\end{itemize}

\section{related work} \label{sec:related_work}

Scalable large-scale geospatial statistical modeling has been
attempted mostly in the univariate ($p = 1$) setting. The authors in 
\cite{goulartparallel, pesquerparallel, rossinisimple, chengaccelerating, tahmasebi2012accelerating} 
worked on parallelizing the predictions
routines using MPI, OpenMP, 
Parallel Virtual Machines (PVMs) and/or Graphics Processing Units (GPUs).
However, their proposals did not include parallelization strategies
for computing the inverse of the covariance matrix, which consumes
approximately 70\% of the execution time. Nevertheless, they expect
a better speedup when the number of locations to be predicted is large.

In parallel kriging literature ($p = 1$),  parallel implementations of the Cholesky
factorization of the kriging process depend on blocking algorithms
and run on a single hardware architecture. For instance,
in \cite{allombert2014out}, a parallel implementation of kriging was developed 
completely on GPU architectures. They proposed
an out-of-core GPU implementation of the kriging process using
a block-based pivoted Cholesky algorithm which is considered more
suitable for GPU compared to CPU. Another example is the parallel
framework in \cite{zhuo2011parallel} which provides end-to-end geospatial
analysis from maximum likelihood estimation to kriging using ScaLAPACK to 
perform distributed-memory implementation on CPUs. However, their
framework does not include any GPU implementation.
Other strategies focused on
replacing the computation of the full covariance matrix with
low-rank approximation methods; see Section~\ref{sec:low_rank} for a review. 

When $p > 1$, the computations become much more challenging. 
In light of this, we introduce a framework to deal with large-scale multivariate 
geospatial statistical modeling that provides both the dense and the TLR-based
approximate versions of MLE operations on very large problem sizes. 
In this work, we depend on the unified software in \cite{abdulah2018parallel} 
which is powered by dense linear algebra task-based algorithms and 
dynamic runtime systems and especially designed for geospatial statistical modeling. 
The software in \cite{abdulah2018parallel} also has an equivalent R 
implementation described in \cite{abdulah2019exageostatr}. 
The framework we propose depends on the asynchronous task-based dense linear
algebra library \chameleon~\cite{chameleon-soft} and the Hierarchical Computations linear algebra library \hicma~\cite{abdulah2019hierarchical}
for support in dense and low-rank matrix computations, respectively.
Both libraries rely on the dynamic runtime system
\starpu~\cite{augonnet2011starpu} to exploit the computing power on shared and distributed-memory systems based on multi-core, many-core, and hybrid hardware architectures.

\section{Overview of the Problem} \label{sec:overview_of_the_problem}
This section describes multivariate spatial modeling and inference based on a given cross-covariance function, with a brief background on low-rank approximation techniques that have been used in the literature to reduce the complexity of log-likelihood function computations.

\subsection{Cross-Covariance Function}
Quantifying spatial dependence of multiple variables is one of the main foci of multivariate geospatial statistical modeling. The key tool for this purpose is the cross-covariance function. It is a matrix-valued function of dimension $p \times p$, parameterized by $\boldsymbol{\theta} \in \mathbb{R}^q$, that describes the degree of dependence between values at two locations $\mathbf{s}_1$ and $\mathbf{s}_2$, and is of the form $\mathbf{C}(\|\mathbf{h}\|; \boldsymbol{\theta})=\left\{C_{ij}(\|\mathbf{h}\|; \boldsymbol{\theta})\right\}_{i,j=1}^{p}$ under the isotropy assumption, where $\mathbf{h} = \mathbf{s}_1 - \mathbf{s}_2 \in \mathbb{R}^d$, $\| \cdot \|$ denotes the Euclidean norm, and $C_{ij}(\|\mathbf{h}\|; \boldsymbol{\theta})=\text{cov}\left\{Z_i(\mathbf{s}_1),Z_j(\mathbf{s}_2) \right\}.$ When $i = j$, $C_{ii}(\|\mathbf{h}\|; \boldsymbol{\theta})$ is called the marginal covariance function and it measures the spatial dependence between the $i$-th variable at $\mathbf{s}_1$ and at $ \mathbf{s}_2$. When $i \neq j$, $C_{ij}(\|\mathbf{h}\|; \boldsymbol{\theta})$ is called the cross-covariance function and it measures the spatial dependence between the $i$-th variable at $\mathbf{s}_1$ and the $j$-th variable at $ \mathbf{s}_2$, for $i,j=1,\ldots,p$. The choice of cross-covariance function $C_{ij}(\|\mathbf{h}\|; \boldsymbol{\theta})$ is data and application-driven. However, $C_{ij}(\|\mathbf{h}\|; \boldsymbol{\theta})$ needs to ensure that the $\boldsymbol{\Sigma}(\boldsymbol{\theta})$ it builds is a positive definite matrix for any $n \in \mathbb{N}$ and any finite set of points $\mathbf{s}_1,\ldots, \mathbf{s}_n $.

\subsection{Mat\'{e}rn Cross-Covariance Function}
The Gaussian geospatial statistical modeling landscape is replete with cross-covariance function models. A comprehensive review on the available models can be found in \cite{genton2015cross}. The parsimonious multivariate Mat\'{e}rn is a popular cross-covariance function \cite{gneiting2010matern, apanasovich2012valid} and has the form
\begin{equation}\label{eqn:multivariate_matern}
C_{ij}(\|\mathbf{h}\|; \boldsymbol{\theta}) = \frac{\rho_{ij} \sigma_{ii} \sigma_{jj}}{2^{\nu_{ij} - 1} \Gamma \left( \nu_{ij} \right)} \left(\frac{\|\mathbf{h}\|}{a}\right)^{\nu_{ij}} \mathcal{K}_{\nu_{ij}}\left(\frac{\|\mathbf{h}\|}{a}\right), 
\end{equation}
for $i,j = 1, \ldots, p$, where $\mathcal{K}_{\nu}(\cdot)$ is the modified Bessel function of the second kind of order $\nu$ and $\Gamma(\cdot)$ is the gamma function. Here $\boldsymbol{\theta}$ includes, for $i=j$, the marginal variance ($\sigma_{ii}^2>0$), smoothness ($\nu_{ii}>0$), and spatial range ($a>0$) parameters, and for $i \neq j$, the colocated correlation ($\rho_{ij}$) and cross smoothness $(\nu_{ij}>0)$ parameters, such that $\nu_{ij} = \frac{1}{2} (\nu_{ii} + \nu_{jj})$ and 
\begin{equation*}
\rho_{ij} = \beta_{ij} \frac{\Gamma(\nu_{ii} + \frac{d}{2})^{1/2}}{\Gamma(\nu_{ii})^{1/2}} \frac{\Gamma(\nu_{jj} + \frac{d}{2})^{1/2}}{\Gamma(\nu_{jj})^{1/2}} \frac{\Gamma\left\{ \frac{1}{2}(\nu_{ii} + \nu_{jj}) \right\}^{1/2}}{\Gamma\left\{ \frac{1}{2}(\nu_{ii} + \nu_{jj}) + \frac{d}{2} \right\}},
\end{equation*}
for any $\sigma_{ii}^2, a, \nu_{ii} > 0$, $d \geq 1$. Here $(\beta_{ij})_{i, j = 1}^{p}$ is a symmetric
and positive definite correlation matrix. 

The parameter $\rho_{ij}$ describes the dependence or correlation between the $i$-th and $j$-th components situated at the same location, through the latent parameter $\beta_{ij}$. When $\beta_{ij} = 0$, $Z_i(\mathbf{s})$
and $Z_j(\mathbf{s})$ are independent. Otherwise, when $\beta_{ij} > 0$ ($\beta_{ij} < 0$), the
two are positively (negatively) dependent. Sample realizations from the parsimonious bivariate Mat\'e{r}n model above are shown in Fig. \ref{fig:simulated_Z} where $\boldsymbol{\theta} = (\sigma_{11}^2, \sigma_{22}^2, a, \nu_{11}, \nu_{22}, \beta)^{\top} = (1, 1, 0.2, 0.5, 1, 0.5)^{\top}$.

\subsection{Multivariate Prediction}
The cross-covariance matrix $\boldsymbol{\Sigma}(\boldsymbol{\theta})$ is crucial
in obtaining predictions of unknown variables at a prediction location and measuring the uncertainty of these predictions. A prediction location may have all or some variables that are missing. The first case happens when there are locations with sensors that collect measurements for atmospheric variables like temperature, precipitation, and wind speed, for example, and one might be interested in predicting the values of these variables at locations with no sensors. The second case occurs when measurements of one variable are difficult or expensive to obtain while measurements of another variable, correlated with the first one, are easy to collect. In this scenario, there will be more locations with data collection instruments for the cheaper variable, while observations will be sparse for the expensive one. The first case is more prevalent in environmental applications wherein sensors measuring different variables simultaneously  were deployed at predetermined sites. Hence, in this work, we assume that all prediction locations are missing the measurements for all $p$ variables. Multivariate geospatial prediction proceeds as follows. Suppose $\mathbf{s}_0 \in \mathbb{R}^d$ is a prediction
location with an unknown vector of $p$ variables $\mathbf{Z}(\mathbf{s}_0)$. Under the squared-error
loss criterion, the best linear unbiased predictor of $\mathbf{Z}(\mathbf{s}_0)$, given 
$\mathbf{Z} = \left\{\mathbf{Z}(\mathbf{s}_1)^{\top}, \ldots, \mathbf{Z}(\mathbf{s}_{n})^{\top} \right\}^{\top},$ also called the cokriging predictor, is 
\begin{equation} \label{eqn:kriging}
\widehat{\mathbf{Z}}(\mathbf{s}_0) = \mathbf{c}_0^{\top} \boldsymbol{\Sigma}(\boldsymbol{\theta})^{-1} \mathbf{Z}.
\end{equation} 
Here $\mathbf{c}_0$ is the $pn \times p$ matrix formed by taking the cross-covariance between $\mathbf{Z}(\mathbf{s}_0)$ and $\mathbf{Z}(\mathbf{s}_r)$, at all sampled locations $\mathbf{s}_r$, $r = 1, \ldots, n$, i.e., 
\begin{equation}
\label{eqn:cokriging_mat}
\mathbf{c}_0 = \left\{ \mathbf{C}(\mathbf{s}_0 - \mathbf{s}_1; \boldsymbol{\theta}), \ldots, \mathbf{C}(\mathbf{s}_0 - \mathbf{s}_{n}; \boldsymbol{\theta}) \right\}^{\top}.
\end{equation}

\subsection{Low-Rank Approximation} \label{sec:low_rank}
Gaussian geospatial statistical modeling relies heavily on the operations done on $\boldsymbol{\Sigma}(\boldsymbol{\theta})$. In the early stages of modeling, $\boldsymbol{\Sigma}(\boldsymbol{\theta})$ has to be formed by evaluating the cross-covariance function at $n$ locations, for all $p$ variables. Parameters then have to be estimated, with the cross-covariance function evaluated every time new sets of parameters are assumed. Further, the Gaussian log-likelihood in Equation~(\ref{eqn:loglikelihood_function}) requires the inverse and the determinant of $\boldsymbol{\Sigma}(\boldsymbol{\theta})$. Prediction also involves the inverse of $\boldsymbol{\Sigma}(\boldsymbol{\theta})$. 

Several techniques to bypass these computing obstacles by exploiting data sparsity have been proposed. Low-rank approximations have been widely used for data modeling. Several low-rank representations of the original Gaussian processes had been proposed during recent years, including predictive process~\cite{banerjee2008gaussian}, where a select set of knots is used to approximate the original process and a low-rank model is obtained. Later, the modeling approaches were extended to multi-resolution approximation~\cite{katzfuss2017multi} to capture spatial structures from different scales. A low-rank approximation can also be applied to Vecchia's representation for the composite likelihood~\cite{huang2018hierarchical}, resulting in reduced computational burden in obtaining the composite likelihood. Moreover, the low-rank approximation can also be applied to cross-covariance matrices with limited loss of information. Another possibility to facilitate computations for exascale modeling is to reduce the complexity of $\boldsymbol{\Sigma}(\boldsymbol{\theta})$ directly. Covariance tapering~\cite{furrer2006covariance}, for example, forces $\boldsymbol{\Sigma}(\boldsymbol{\theta})$ to be sparse by introducing a compact support. This technique is also known as the Diagonal Super Tile (DST) wherein the entries of the tiles that are very far from the diagonal are annihilated or reduced to zero \cite{abdulah2018exageostat}. Bayesian hierarchical models have also been championed in large-scale geospatial statistical analyses \cite{rue2009approximate, banerjee2017high}.

\subsection{Univariate MLOE/MMOM}

A common metric used to assess the quality of the predictions is the MSPE and is computed as $  \text{MSPE} = \frac{1}{n_{\text{pred}}} \sum_{l = 1}^{n_{\text{pred}}} \| \widehat{\mathbf{Z}}(\mathbf{s}_{0,l}) - \mathbf{Z}(\mathbf{s}_{0,l}) \|^2,$
where $\mathbf{s}_{0,1}, \mathbf{s}_{0,2}, \ldots, \mathbf{s}_{0,n_{\text{pred}}}$ are the $n_{\text{pred}}$ prediction locations.
When the predictions are obtained through an approximated cross-covariance matrix, e.g., the TLR version of the exact covariance matrix, an appropriate metric should be used. The authors in \cite{hong2019efficiency} suggested the use of the MLOE and MMOM. However, their formulations are only available for univariate predictions ($p = 1$). The formulas and the algorithm for the univariate MLOE/MMOM can be found in \cite{hong2019efficiency}. In Section \ref{sec:proposed_bivariate_mloe_mmom}, we extend these metrics and the algorithm to multivariate. 

\section{Proposed Framework} \label{sec:proposed_framework}

In this section, we explain in detail our software dependencies, proposed multivariate modeling and inference implementation, and multivariate MLOE/MMOM criteria which are used to assess the accuracy of the multivariate modeling.

\begin{figure}[t!]
    \centering
    \includegraphics[width=0.37\textwidth]{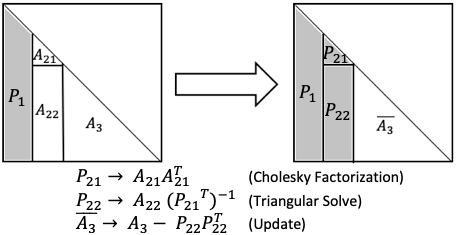}
    \caption{One Shot of the Block Cholesky Factorization Algorithm.}
    \label{fig:fact_block_alg}
\end{figure}

\subsection{Parallel Software Architecture} \label{sec:software}
\begin{figure}[t!]
    \centering
    \includegraphics[width=0.4\textwidth]{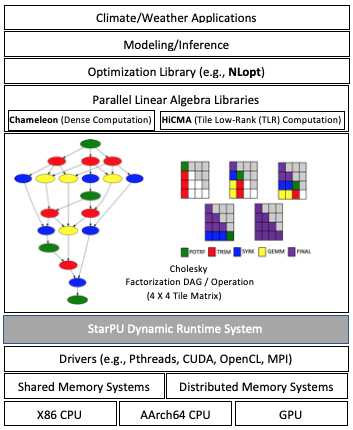}
    \caption{Software architecture based on \nlopt, \chameleon, \hicma,  and \starpu for climate/weather applications with a $4 \times 4$  Cholesky factorization DAG example.}
    \label{fig:starpu_graph}
\end{figure}

Our proposed framework internally relies on a list of software dependencies
including \chameleon~\cite{chameleon-soft}, \starpu~\cite{augonnet2011starpu},
\hicma~\cite{abdulah2019hierarchical}, \starsh~\cite{starsh-soft}, and 
\nlopt~\cite{johnson2014nlopt}, as demonstrated in Fig.~\ref{fig:starpu_graph}. 
\chameleon is a tile-based high-performance numerical library based on task-parallel 
programming models which offer a more structured way of expressing parallelism
using three backend runtime systems, namely, QUARK~\cite{yarkhan2011quark}, 
\parsec~\cite{bosilca2013parsec}, and \starpu. 

In the literature, parallel linear algebra operations were performed in 
parallel systems using block-column algorithms. These
algorithms disband the given matrix, represented in column-major format, 
into a successive panel and update the computational phases. Assuming that 
Cholesky factorization is performed, the factorization
is applied to each panel, and the matrix transformations are blocked 
and applied together at one time during the update phase; see Fig.~\ref{fig:fact_block_alg}. LAPACK
and ScaLAPACK are examples of dense linear algebra library on shared-memory
 and distributed-memory systems, respectively.
\chameleon  adopts tile algorithms methodology by splitting the given matrix into a 
set of tiles, instead of panels, which allows updating  the trailing submatrix
before factorization is complete. These algorithms aim at weakening the synchronization points
while performing matrix operations and maximization of the utilization of underlying hardware resources.
The numerical algorithm can then be
translated into a Directed Acyclic Graph (DAG), where the
nodes represent tasks and the edges define data dependencies (e.g., read, write, and read-write), 
as shown in Fig.~\ref{fig:starpu_graph} 
with a $4 \times 4$ Cholesky factorization DAG example.
Runtime systems lead the rudder of utilizing the usage of underlying
hardware resources in \chameleon, allowing tile algorithms to run efficiently on different parallel
hardware with homogeneous and heterogeneous architectures. DAG tasks are scheduled across
different hardware resources to ensure that the data dependencies rules predefined by the user
are not violated. Runtime systems enhance software productivity by abstracting the hardware
complexity from the end-user. They are also
capable of mitigating data movement overhead, reducing load imbalance, and increasing hardware utilization.

Here, we use the \starpu
dynamic runtime system because of its ability to support a wide range of parallel
heterogeneous hardware architectures from different vendors like Intel, AMD,
NVIDIA, and ARM. \starpu  executes defined generic task graphs, generated
by a built-in sequential task flow (STF) programming model. \starpu scheduler pushes the set of tasks to the
available processing unit based on these dependencies which may lead to asynchronous
execution. \starpu supports different programming languages (e.g., Pthreads, CUDA, OpenCL, and MPI) 
and runs on different hardware architectures (e.g., CPU/GPU, shared/distributed-memory).
\starpu may decide at runtime to execute the tasks on different hardware based on performance models.

\hicma supports parallel TLR matrix computation. It
relies on \chameleon with \starpu as the runtime and \starsh
as the compressed matrix generator. \nlopt is an open-source C/C++ nonlinear
optimization toolbox which we rely on to perform the optimization task for the
MLE operation. 
\starpu runtime system handles both exact dense and TLR
computational workloads to perform the required linear algebra operations in parallel.

\subsection{Exact Multivariate Modeling} \label{sec:exact_multivariate_modeling}
\begin{figure}[t!]
    \centering
    \includegraphics[width=0.4\textwidth]{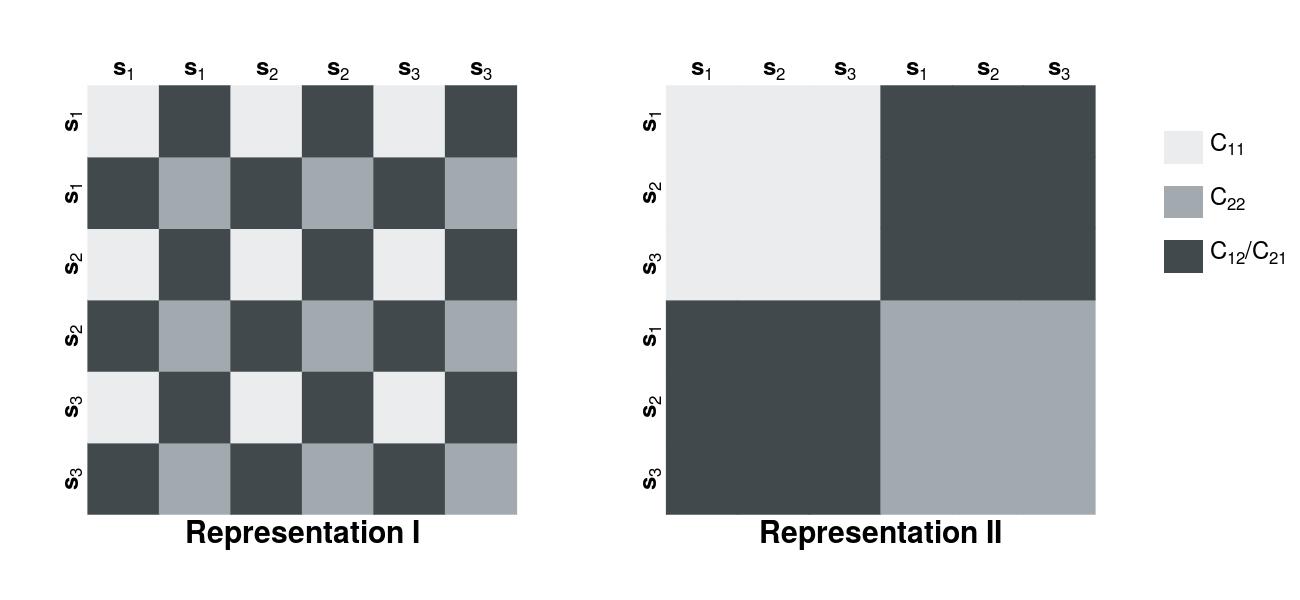}
    \caption{The order of the elements of $\boldsymbol{\Sigma}(\boldsymbol{\theta})$ is drawn under the two different cross-covariance matrix representations.}
    \label{fig:matrix_representation}
\end{figure}

In multivariate modeling, there are two ways to build $\boldsymbol{\Sigma}(\boldsymbol{\theta})$~\cite{genton2015cross}. The first approach (Representation I) is to build an $n\times n$ matrix with block elements of $p\times p$ matrices $\mathbf{C}(\mathbf{s}_{l}-\mathbf{s}_{r}; \boldsymbol{\theta})$, $l,r = 1,\ldots,n$. The second approach (Representation II) is to build a $p\times p$ matrix with block elements of $n\times n$ matrices $\left\{C_{ij}(\mathbf{s}_{l}-\mathbf{s}_{r}; \boldsymbol{\theta})\right\}_{l, r = 1}^{n}$, $i, j = 1, \ldots, p$. To illustrate, suppose $p = 2$ and $n = 3$. Fig.~\ref{fig:matrix_representation} shows $\boldsymbol{\Sigma}(\boldsymbol{\theta})$ of dimension $6 \times 6$ under the two representations.

The two representations in Fig.~\ref{fig:matrix_representation} yield a symmetric positive definite matrix. Cholesky factorization, the backbone of MLE, is performed on this symmetric positive definite matrix to obtain its inverse and log determinant required for maximizing the log-likelihood function in Equation~(\ref{eqn:loglikelihood_function}).

A simulation study on the parsimonious bivariate Mat\'{e}rn was conducted to assess the efficiency in parameter estimation and accuracy of predictions under the two representations via comparison of the medians and standard deviations of the estimated parameters and the MSPE of the predictions.  The results indicate that the two representations are numerically equivalent in exact computation and either one can substitute for the other. Thus, only Representation I of the exact multivariate model is utilized in this work. 

To hasten parameter estimation of the exact parsimonious Mat\'{e}rn cross-covariance function, we maximize the profile log-likelihood in lieu of the full log-likelihood in Equation (\ref{eqn:loglikelihood_function}). The profile likelihood is a variant of the full log-likelihood wherein the number of parameters to be estimated is effectively reduced \cite{severini2000likelihood}. Under this approach, the marginal variance parameters $\mathbf{\sigma}_{ii}^2$, for $i = 1, \ldots, p$, are no longer included in the estimation and can be derived after all the other parameters were estimated, i.e., $\hat{\sigma}_{ii}^2 = n^{-1} \{ \mathbf{Z}_{i}^{\top} \mathbf{R}_{ii}(\hat{\boldsymbol{\theta}}_i)^{-1} \mathbf{Z}_{i} \} $, where $\hat{\boldsymbol{\theta}}_i \in \mathbb{R}^{q_i}$, $q_i \geq 1$, is the set of $q_i$ estimated marginal parameters for $Z_i$, except the marginal variance parameter $\sigma_{ii}^2$, i.e., $\hat{\nu}_{ii}, \hat{a}$, and $\mathbf{R}_{ii}(\hat{\boldsymbol{\theta}}_i)$ is the correlation matrix formed by evaluating the cross-covariance function for $i = j$, using $\hat{\boldsymbol{\theta}}_i$ and $\sigma_{ii}^2 = 1$, for $i = 1, \ldots, p$. Here $\mathbf{Z}_i$ is the vector formed by all the values pertaining to variable $i$, i.e., $\mathbf{Z}_i = \left\{Z_i(\mathbf{s}_1), \ldots, Z_i(\mathbf{s}_n)\right\}^{\top}$. 

\subsection{TLR-Based Multivariate Modeling}

Through the last decade, tile algorithms were created to adapt to parallel architectures that require data sharing~\cite{buttari2009class}. The tiling mechanism improved block-based algorithm which
suffers from the existence of numerous synchronization points that slow down the overall performance.

\begin{figure}[t!]
    \centering
    \includegraphics[width=0.32\textwidth]{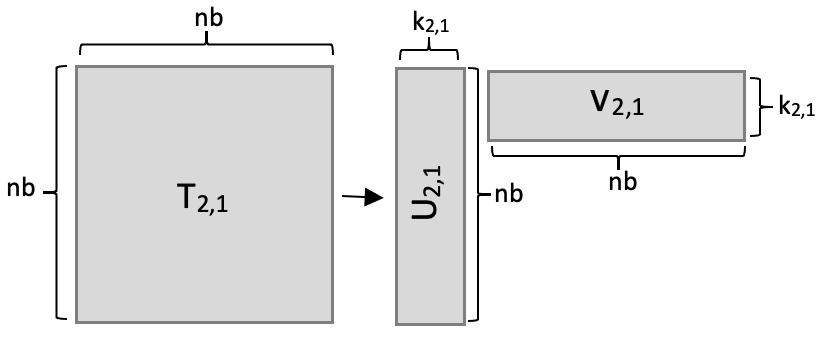}
    \caption{An example of TLR approximation tile: $T_{2, 1}$ with dimension $nb \times nb$ is approximated by two matrices $U_{2, 1}$ and $V_{2, 1}$ with dimension $nb \times k_{2, 1}$}
    \label{fig:tlr-example}
\end{figure}

TLR approximation algorithms have been implemented based
on the tiling technique. Instead of applying the low-rank approximation to the whole
matrix, each tile is compressed as a separate unit. Here, we rely on the 
TLR implementation of~\cite{akbudak2018exploiting}, where the authors have implemented 
the TLR approximation by performing the singular value decomposition (SVD) algorithm
for each off-diagonal tile by preserving the most significant values and vectors
in the corresponding tile, i.e., the tile rank. The diagonal tiles are kept dense since
they cannot be approximated. Ranks of the off-diagonal tiles are determined based on the accuracy requirement of the application. Fig. \ref{fig:tlr-example} shows an example of compressing an off-diagonal tile $T_{2, 1}$ to two matrices $U_{2, 1}$ and $V_{2,1}$.

\begin{figure}
\centering
\begin{minipage}[h]{.5\textwidth}
\includegraphics[height=1.2in]{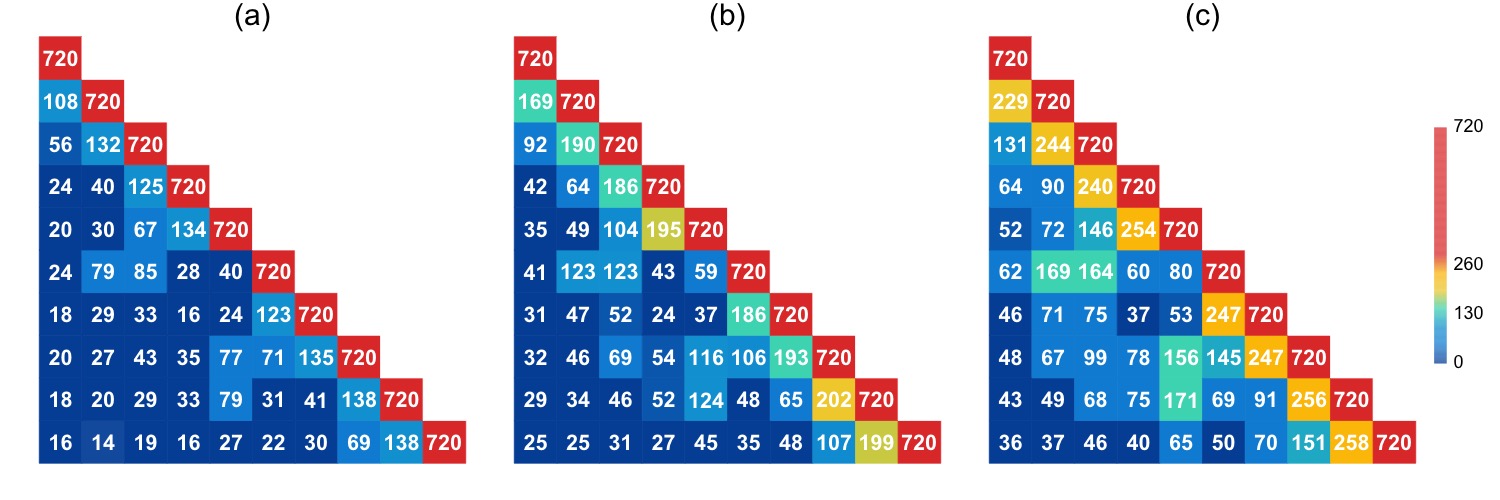}
\caption{Rank distributions of a $7200 \times 7200$ bivariate covariance matrix using $nb$ = 720 with parsimonious bivariate Mat\'{e}rn parameters $\boldsymbol{\theta} = (1, 1, 0.09, 0.5, 1, 0.5)^{\top}$ under (a) TLR5, (b) TLR7, and (c) TLR9. }
\label{fig:tlr-ranks}
\end{minipage}\qquad \qquad \qquad
\end{figure}

The effectiveness of the TLR mechanism depends on the ranks of the matrix tiles, which in turn depend on the accuracy requirements of the given application. To reduce the ranks of the covariance matrix tiles, we ordered the matrix based on the Morton ordering approach~\cite{morton-zorder-1966}, which matches with Representation I in Fig.~\ref{fig:matrix_representation}. To validate the usage of the
TLR approximation with multivariate modeling, we estimate the ranks of the generated covariance
matrix tiles with different accuracy levels, namely, TLR5 ($10^{-5}$), TLR7 (1$0^{-7}$), and TLR9 ($10^{-9}$), on a $7200 \times 7200$ bivariate
covariance matrix; see Fig.~\ref{fig:tlr-ranks}. The ranks distribution shows that the off-diagonal ranks grow as the tiles get closer to the diagonal. It can also be observed that even with a higher accuracy, e.g., TLR9, the ranks are still small compared to the ranks of the full dense tiles in the diagonal (in red). The example was drawn from a synthetic set of parameters $\boldsymbol{\theta} = (\sigma_{11}^2, \sigma_{22}^2, a, \nu_{11}, \nu_{22}, \beta)^{\top} = (1, 1, 0.09, 0.5, 1, 0.5)^{\top}$, which represents a moderate spatial dependence between two variables, $Z_1$ and $Z_2$. Other sets of parameters representing varying strengths of spatial dependence were also examined and the results do not differ significantly from what is shown in Fig.~\ref{fig:tlr-ranks}. 

\begin{figure}[t!]
    \centering
    \includegraphics[width=0.32\textwidth]{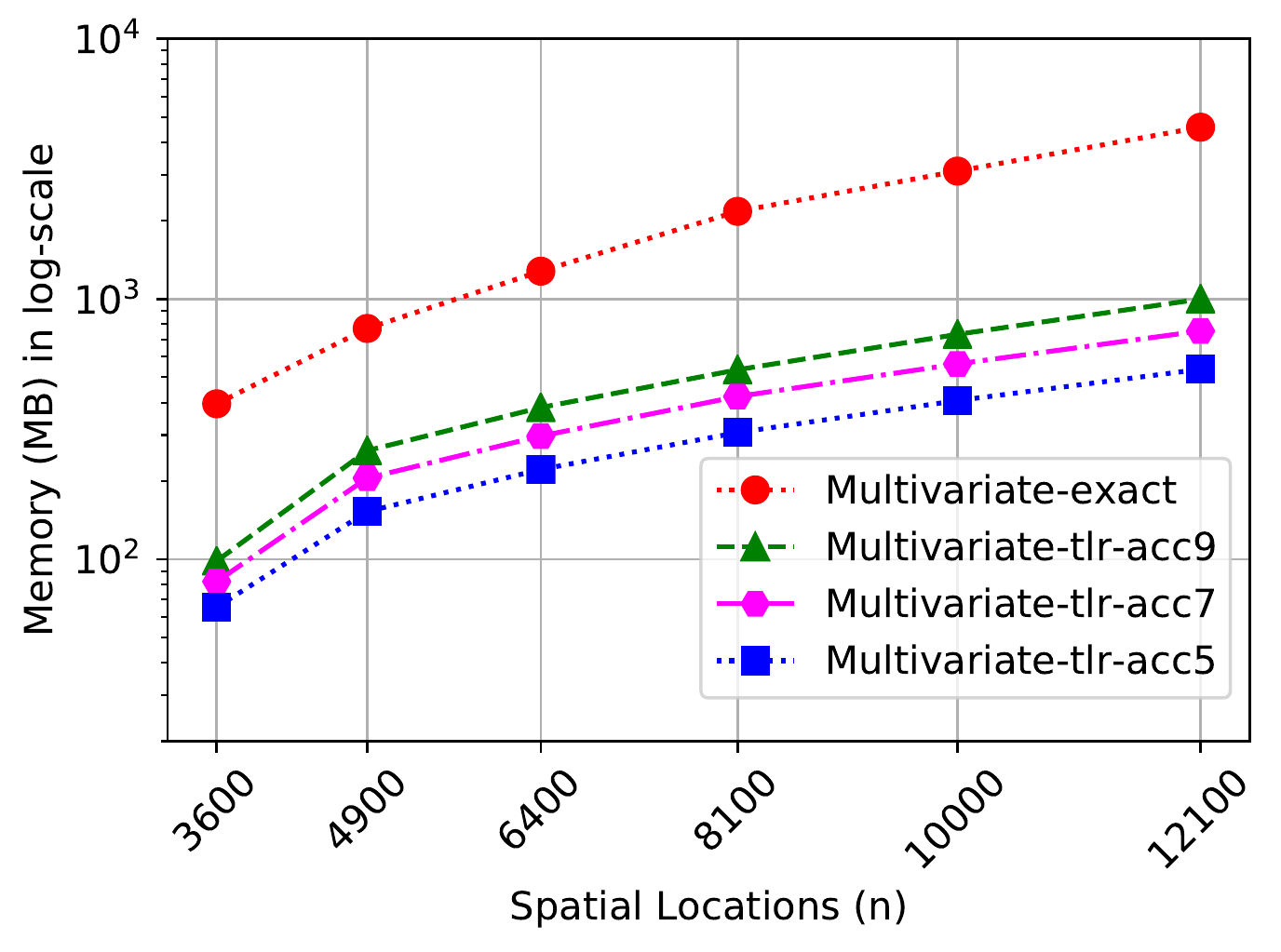}
    \caption{Memory footprint of  exact and TLR-based MLE with varying size of $n$ and two measurement vectors $Z_1$ and $Z_2$.}
    \label{fig:mem-footprint-example}
\end{figure}

Fig.~\ref{fig:mem-footprint-example} shows the memory footprint of
the requirement of full MLE operation for dense, TLR5, TLR7, and TLR9
on various multivariate problem sizes. The memory footprint involves
$\boldsymbol{\Sigma}(\boldsymbol{\theta})$ of dimension $2n \times 2n$ 
and two measurement vectors, $Z_1$ and  $Z_2$. 
The measurement vectors are  always represented in the exact format since 
there is no benefit in compressing them. As shown, the TLR-based 
compression requires less memory footprint with
respect to the dense representation. On average, the TLR representations
require 6.68X, 4.93X, and 3.86X less memory than the dense
representation for $10^{-5}$, $10^{-7}$, and $10^{-9}$, respectively.
The memory saving increases with larger problem sizes, as seen in the figure.

The performance model of TLR is driven by the most time-consuming kernel, i.e., the TLR matrix-matrix multiplication (TLR-MM).
The arithmetic complexity of a single TLR-MM is $36 \times nb \times k^2$~\cite{pasc2020}, with $nb$ the tile size and $k$ the tile rank 
that depends on the number of significant singular values after compression. 
The total number of operations is $O(n^2 k)$, attained when 
$nb = O(\sqrt{n})$. This tile size is a trade-off between the arithmetic intensity of the kernel and the degree of parallelism
of the algorithm. The performance model of TLR is driven by a quadratic regime, which contrasts with the cubic regime 
for exact computations. 
The detailed complexity analysis of TLR Cholesky factorization can be found in~\cite{Mary2018}.
We are also looking into more advanced matrix compression strategies~\cite{Ambikasaran2013,AminfarD14,hackbusch2015hierarchical} 
that exhibit better arithmetic complexity but these may be challenging to implement on massively parallel systems 
due to their hierarchical structures. 

\subsection{Proposed Multivariate MLOE/MMOM Criteria} \label{sec:proposed_bivariate_mloe_mmom}

Assessing the estimation accuracy of the modeling approach is challenging and requires
a well-developed algorithm. Our novel multivariate prediction assessment metrics depend on the MLOE/MMOM~\cite{hong2019efficiency}.  There are two possible multivariate versions of the MLOE/MMOM. The first version is a naive extension and it is simply the mean of the MLOE/MMOMs of all the variables. This approach requires univariate covariance models and the univariate version of the cokriging equation~(\ref{eqn:kriging}).
Another version, which we propose, is to utilize cross-covariance functions and the cokriging equation (\ref{eqn:kriging}). Denote the cokriging error vector by $\mathbf{e} (\mathbf{s}_{0}) = \left(e_{1}, \ldots, e_{p} \right)^{\top}$, where $e_{i} = \hat{Z}_i(\mathbf{s}_0) - Z_i(\mathbf{s}_0)$ and $\hat{Z}_i (\mathbf{s}_0)$ is the predictor for variable $i$, $i = 1, \ldots, p$, at a prediction location $\mathbf{s}_0$, obtained using the cokriging equation~(\ref{eqn:kriging}), with the true cross-covariance function parameters, $\boldsymbol{\theta}$. The mean square error of this predictor is
\begin{equation}\label{mse_predictor}
\text{E}_{t} = \text{tr} \left\{\mathbf{C} (\mathbf{0}; \boldsymbol{\theta}) - {\mathbf{c}_0^{t}}^{\top} \boldsymbol{\Sigma} (\boldsymbol{\theta})^{-1} \mathbf{c}_0^{t} \right\},
\end{equation}
where the subscript $t$ in $\mathbf{c}_0^t$ specifies that the parameters used in Equation (\ref{eqn:cokriging_mat}) are the true parameters $\boldsymbol{\theta}$ and tr indicates the trace of the matrix. Suppose now that the set of estimated parameters derived from using a certain approximation of the covariance matrix, $\hat{\boldsymbol{\theta}}^a$, were used to build the cokriging equation (\ref{eqn:kriging}). The error introduced by the approximation is $\mathbf{e}_{a} (\mathbf{s}_{0}) = \left(e_{a,1}, \ldots, e_{a,p} \right)^{\top}$, where $e_{a,i} = \hat{Z}_i^a(\mathbf{s}_0) - Z_i(\mathbf{s}_0)$ and $\hat{Z}_i^a (\mathbf{s}_0)$ is the predictor for variable $i$ at a prediction location $\mathbf{s}_0$, obtained using the cokriging equation in (\ref{eqn:kriging}) with $\hat{\boldsymbol{\theta}}^a$. The mean square error of this predictor is 
\begin{multline} \label{mse_predictor2}
\text{E}_{t,a}  = \text{tr} \big\{ \mathbf{C}(\mathbf{0}; \boldsymbol{\theta}) - 2{\mathbf{c}_0^{t}}^{\top} \boldsymbol{\Sigma}  (\hat{\boldsymbol{\theta}}^a)^{-1} \mathbf{c}_0^{a} \\
+ {\mathbf{c}_0^{a}}^{\top} \boldsymbol{\Sigma} (\hat{\boldsymbol{\theta}}^a)^{-1}\boldsymbol{\Sigma} (\boldsymbol{\theta}) \boldsymbol{\Sigma}  (\hat{\boldsymbol{\theta}}^a)^{-1} \mathbf{c}_0^{a} \big\}
\end{multline}
where $\mathbf{c}_0^{a}$ is Equation (\ref{eqn:cokriging_mat}) evaluated using $\hat{\boldsymbol{\theta}}^a$ and $\boldsymbol{\Sigma} (\hat{\boldsymbol{\theta}}^a)$ is the cross-covariance matrix also evaluated using $\hat{\boldsymbol{\theta}}^a$. The subscript $t,a$ in $\text{E}_{t,a}$ specifies that given the true parameters $\boldsymbol{\theta}$, the estimated parameters $\hat{\boldsymbol{\theta}}^a$ from the approximated model are used instead.

The multivariate MLOE/MMOM, denoted as $\text{MLOE}^{CK}$ and $\text{MLOE}^{CK}$, respectively, are as follows:
\begin{equation} \label{eqn:mloe_multi}
 \setlength{\thickmuskip}{.1\thickmuskip}
  \setlength{\medmuskip}{.1\medmuskip}
\text{MLOE}^{CK} = \frac{1}{n_{\text{pred}}} \sum_{l = 1}^{n_{\text{pred}}} \text{LOE}^{CK}(\mathbf{s}_{0,l})
\end{equation}
\begin{equation} \label{eqn:mmom_multi}
 \setlength{\thickmuskip}{.0001\medmuskip}
  \setlength{\medmuskip}{.0001\medmuskip}
\text{MMOM}^{CK} = \frac{1}{n_{\text{pred}}} \sum_{l = 1}^{n_{\text{pred}}} \text{MOM}^{CK}(\mathbf{s}_{0,l}),
\end{equation}
where $\text{LOE}^{CK}(\mathbf{s}_{0}) = \frac{\text{E}_{t,a}}{\text{E}_{t}} - 1$ and $\text{MOM}^{CK}(\mathbf{s}_{0}) = \frac{\text{E}_{a}}{\text{E}_{t,a}} - 1$. The superscript $CK$ stipulates that the multivariate extensions were derived from the cokriging equation (\ref{eqn:kriging}) and $\text{E}_a$ in Equation (\ref{mse_predictor}) evaluated using $\hat{\boldsymbol{\theta}}^a$ and $\mathbf{c}_0^{a}$.

The algorithm implementing this approach for $p = 2$ is outlined in Algorithm~\ref{alg:multivariate_mloe_mmom}. This new algorithm is similar to~\cite{hong2019efficiency} except now the matrix-valued Mat\'{e}rn cross-covariance function is utilized instead of the scalar-valued univariate Mat\'{e}rn covariance function, e.g., $\text{BiMat\'{e}rn}(\mathbf{s}_1, \mathbf{s}_2; \boldsymbol{\theta})$ returns a $2 \times 2$ matrix. 
\begin{center}
\begin{algorithm}
       \caption{\small{Algorithm for Parallel Bivariate MLOE/MMOM.}} 
        \scalebox{0.74}{
 \vbox{ 
     \begin{minipage}{11.5cm}
  \begin{algorithmic}[1]
 \renewcommand{\algorithmicrequire}{\textbf{Result:} }
\renewcommand{\algorithmicensure}{\textbf{Input:}}
 \ENSURE  $n_{\text{pred}}$: number of prediction locations; $n$: number of sampled locations; $\mathbf{s}_{0,1}, \ldots, \mathbf{s}_{0,n_{\text{pred}}}$: 
prediction locations; $\mathbf{s}_{1}, \ldots, \mathbf{s}_{n}$: sampled locations; dist$(\mathbf{a}, \mathbf{b})$: function that computes 
the distance between $\mathbf{a}$ and $\mathbf{b}$; $\boldsymbol{\theta}$: true 
parameters for the bivariate Mat\'{e}rn; $\hat{\boldsymbol{\theta}}^a$: estimated 
parameters for the approximated bivariate Mat\'{e}rn; $\text{BiMat\'{e}rn}(\mathbf{a}, \mathbf{b}; \boldsymbol{\theta})$: bivariate Mat\'{e}rn covariance function evaluated at locations $\mathbf{a}$ and $\mathbf{b}$ (Equation \ref{eqn:multivariate_matern}); $\text{CovMat}(\boldsymbol{\theta})$: $2 n \times 2 n$ cross-covariance matrix;
 \REQUIRE MLOE $:=$ mean of LOE$^{CK}$ , MMOM $:=$ mean of MOM$^{CK}$
 \STATE $\boldsymbol{\Sigma} = \text{CovMat}(\boldsymbol{\theta})$;
  \STATE $\boldsymbol{\Sigma}^{a} = \text{CovMat}( \hat{\boldsymbol{\theta}}^a)$;
 \STATE $\mathbf{L} \mathbf{L}^{\top} = \boldsymbol{\Sigma}$; \COMMENT{cholesky factorization}
   \STATE $\mathbf{L}^a (\mathbf{L}^a)^{\top} = \boldsymbol{\Sigma}^a $; \COMMENT{cholesky factorization}
   
  \FOR {$l = 1$ to $n_{\text{pred}}$}
  \FOR {$r = 1$ to $n$}
  \STATE $\mathbf{c}_0^{t}[(2r - 1):(2r), 1:2] = \text{BiMat\'{e}rn}(\mathbf{s}_r, \mathbf{s}_{0,l}; \boldsymbol{\theta})$; \COMMENT{Eq. (\ref{eqn:cokriging_mat})}
  \STATE $\mathbf{c}_0^{a}[(2r - 1):(2r), 1:2] = \text{BiMat\'{e}rn}(\mathbf{s}_r, \mathbf{s}_{0,l}; \hat{\boldsymbol{\theta}}^a)$; \COMMENT{Eq. (\ref{eqn:cokriging_mat})}
  \ENDFOR
  
  \STATE tmp1 $ = $ $ \text{tr} \big\{ \text{BiMat\'{e}rn}(\mathbf{s}_{0,l}, \mathbf{s}_{0,l}; \boldsymbol{\theta}) - 2{\mathbf{c}_0^{t}}^{\top} (\mathbf{L}^a)^{-\top} (\mathbf{L}^a)^{-1}  \mathbf{c}_0^{a}+ {\mathbf{c}_0^{a}}^{\top} (\mathbf{L}^a)^{-\top} (\mathbf{L}^a)^{-1} \mathbf{L} \mathbf{L}^{\top} (\mathbf{L}^a)^{-\top} (\mathbf{L}^a)^{-1} \mathbf{c}_0^{a} \big\}$; \COMMENT{Eq. (\ref{mse_predictor2})}
  \STATE tmp2 $ = $ $ \text{tr} \left\{ \text{BiMat\'{e}rn}(\mathbf{s}_{0,l}, \mathbf{s}_{0,l}; \boldsymbol{\theta}) - {\mathbf{c}_0^{t}}^{\top} \mathbf{L}^{- \top} \mathbf{L}^{-1} \mathbf{c}_0^{t} \right\}$ ; \COMMENT{Eq. (\ref{mse_predictor})}
  \STATE tmp3 $ = $ $\text{tr}\left\{ \text{BiMat\'{e}rn}(\mathbf{s}_{0,l}, \mathbf{s}_{0,l}; \hat{\boldsymbol{\theta}}^a) - {\mathbf{c}_0^{a}}^{\top} (\mathbf{L}^a)^{-\top} (\mathbf{L}^a)^{-1} \mathbf{c}_0^{a} \right\}$; \COMMENT{Eq. (\ref{mse_predictor})}
  \STATE $\text{LOE}^{CK}[l] = \text{tmp1} / \text{tmp2} - 1;$  \COMMENT{Eq. (\ref{eqn:mloe_multi})}
  \STATE $\text{MOM}^{CK}[l] = \text{tmp3} / \text{tmp1} - 1;$ \COMMENT{Eq. (\ref{eqn:mmom_multi})}
  \ENDFOR
 \end{algorithmic} 
 \end{minipage}}
 }
  \label{alg:multivariate_mloe_mmom}
      \end{algorithm}

\end{center}

Assuming $n_{\text{pred}} \ll n$, the memory footprint and the arithmetic complexity of  Algorithm~\ref{alg:multivariate_mloe_mmom} for any values of $p$ depends solely on the Cholesky factorizations of $\boldsymbol{\Sigma}$ and $\boldsymbol{\Sigma}^a$ (lines 3 and 4). Each Cholesky factorization requires $p^2 n^2$ memory footprint and $(1/3) p^3 n^3$ number of operations. Indeed,
the code section containing the nested loops (lines 5-15) carries only Level-1 and Level-2 BLAS operations involving several dot products and triangular solves. Since these matrix operations account for the lower order terms~\cite{anderson1992lapack}, the overall memory footprint then is $2p^2 n^2$ and the arithmetic complexity is $(2/3) p^3 n^3$.  

\section{Performance Results} \label{sec:performance}
In this section, we assess the performance and accuracy of the TLR approximation to $\boldsymbol{\Sigma}(\boldsymbol{\theta})$ on large-scale simulations and real datasets. The performance assessment involves a wide range of parallel hardware systems while the accuracy assessment entails simulating synthetic datasets from the parsimonious Mat\'{e}rn cross-covariance function, estimating the model parameters, and predicting values at screened locations. Estimation and prediction accuracy are also assessed on real datasets. The designed experiments show that the TLR approximation outperforms the exact bivariate computation while maintaining the accuracy required by geospatial statistics applications.


\subsection{Testbed and Methodology}
\begin{figure*}[t!]
    \centering
    \begin{subfigure}[t]{0.23\textwidth}
        \centering
        \includegraphics[height=1.3in]{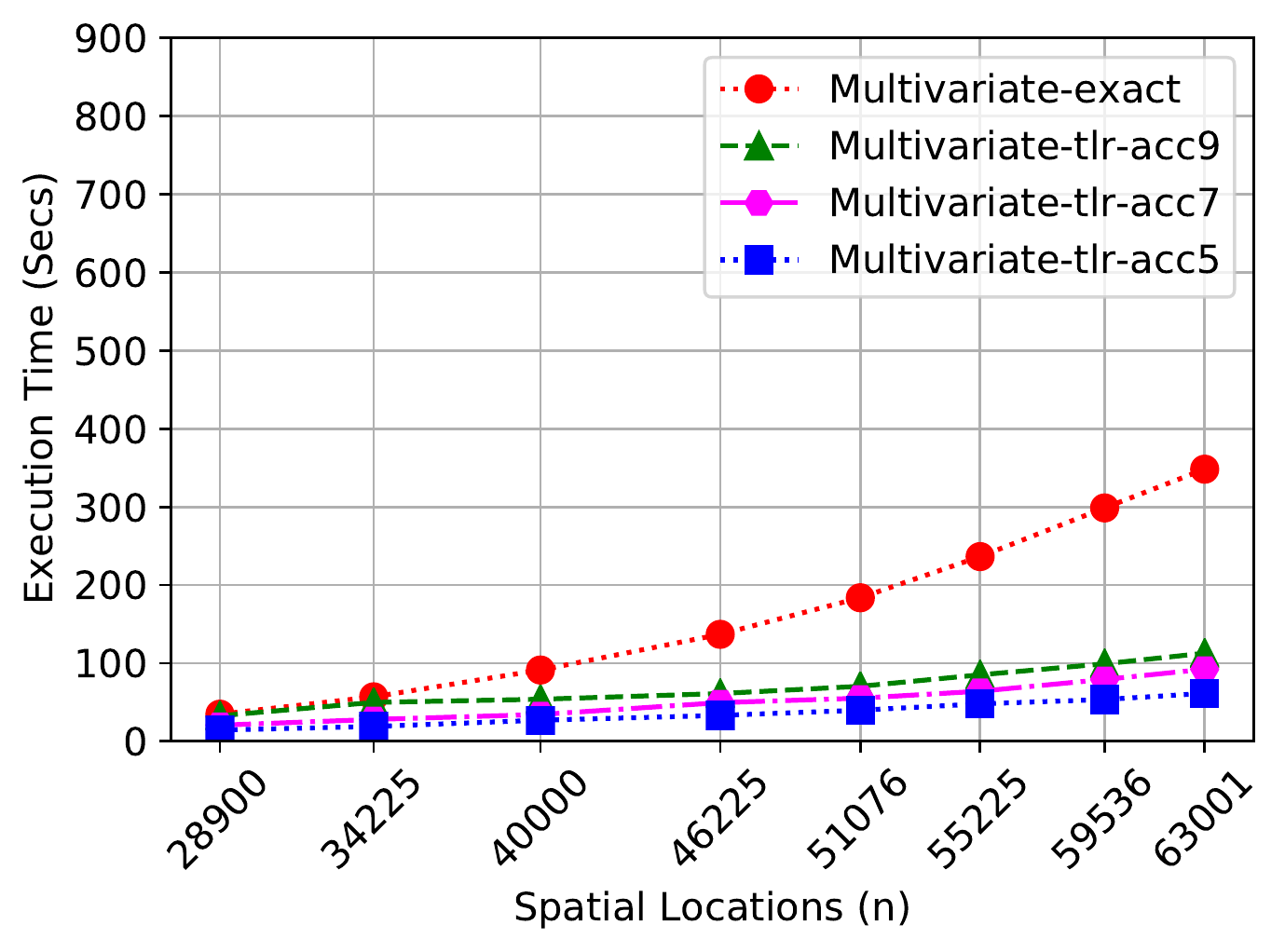}
        \caption{56-core Intel Skylake}
    \end{subfigure}%
    ~ 
    \begin{subfigure}[t]{0.23\textwidth}
        \centering
        \includegraphics[height=1.3in]{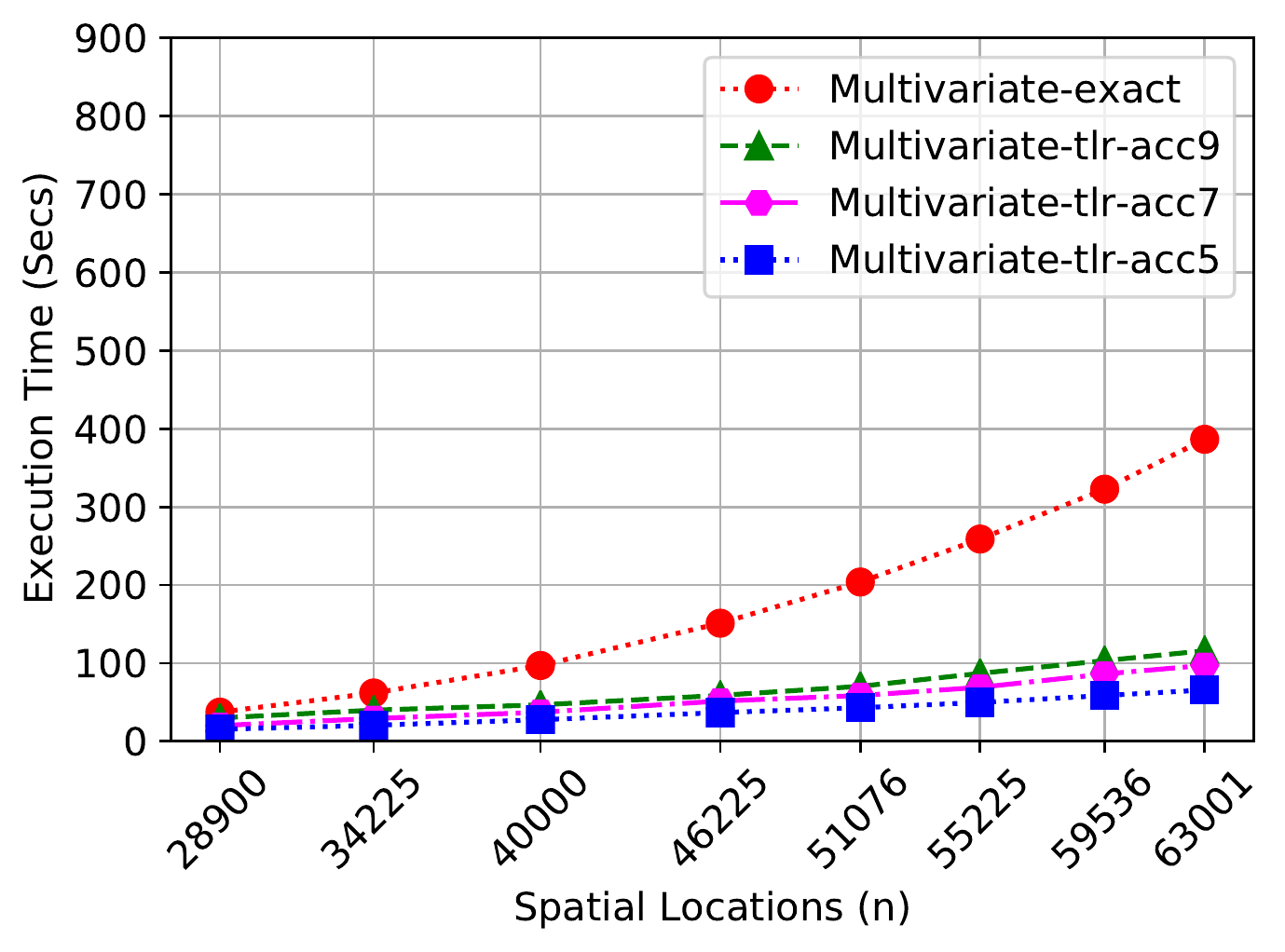}
        \caption{40-core Intel Cascade Lake}
    \end{subfigure}
    ~
        \begin{subfigure}[t]{0.23\textwidth}
        \centering
        \includegraphics[height=1.3in]{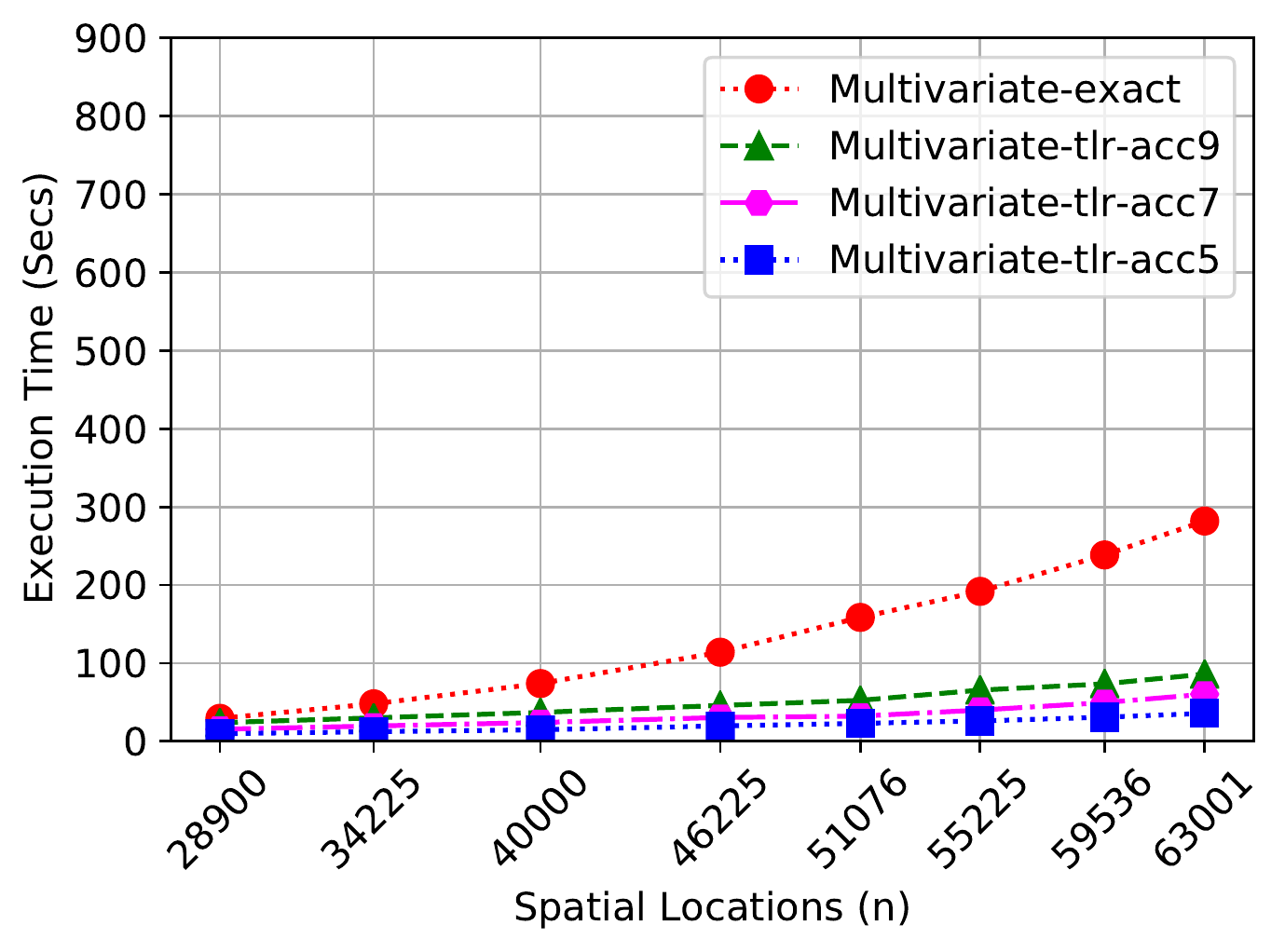}
        \caption{128-core AMD EPYC}
    \end{subfigure}
    ~
            \begin{subfigure}[t]{0.23\textwidth}
        \centering
        \includegraphics[height=1.3in]{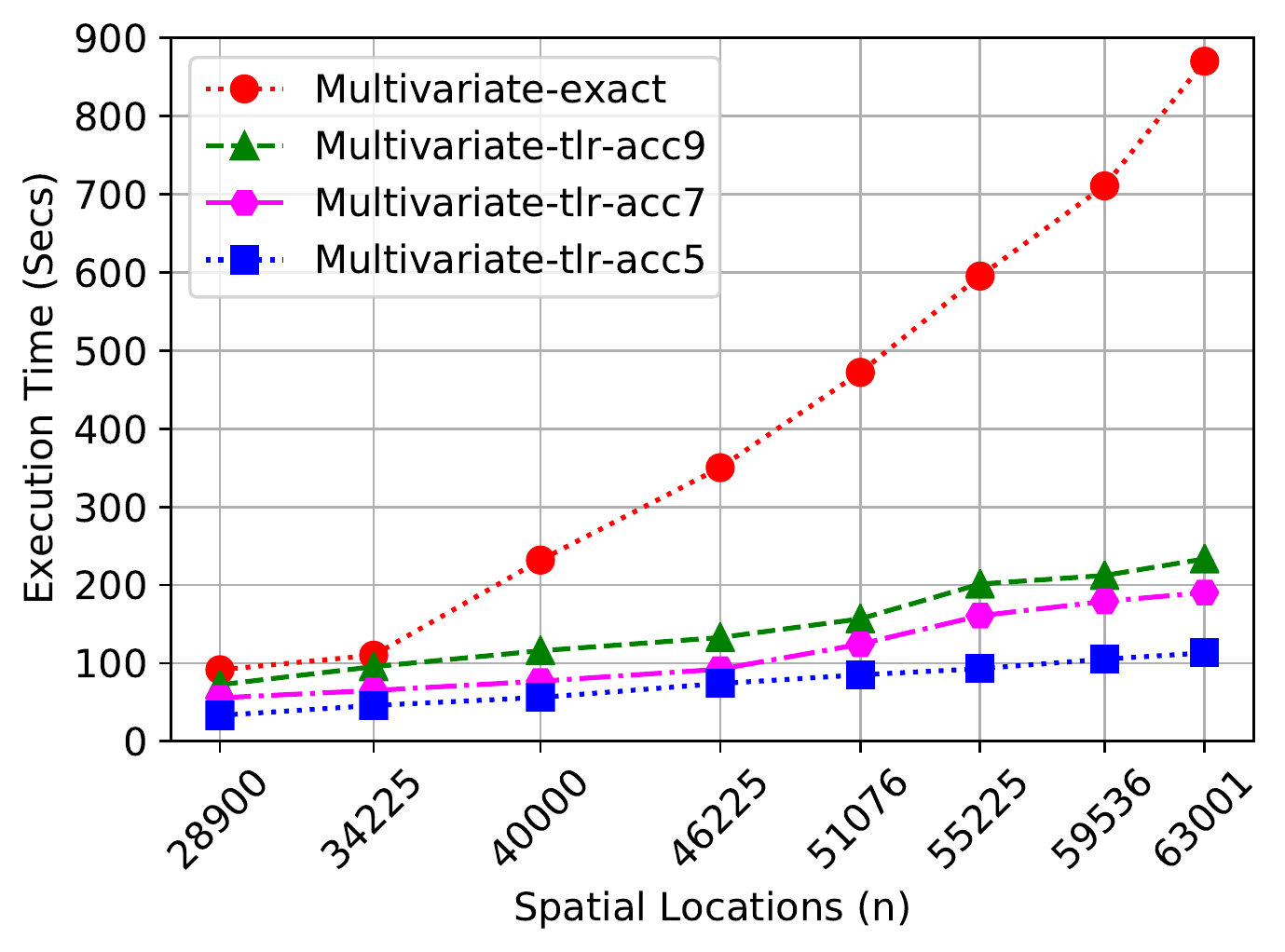}
        \caption{64-core ThunderX2 ARM}
    \end{subfigure}

    \caption{One bivariate exact and TLR-based MLE iteration  using different matrix sizes on various shared-memory systems. }
        \label{fig:tlr}
\end{figure*}

\begin{figure}[t!]
    \centering
    \begin{subfigure}[t]{0.23\textwidth}
        \centering
        \includegraphics[height=1.25in]{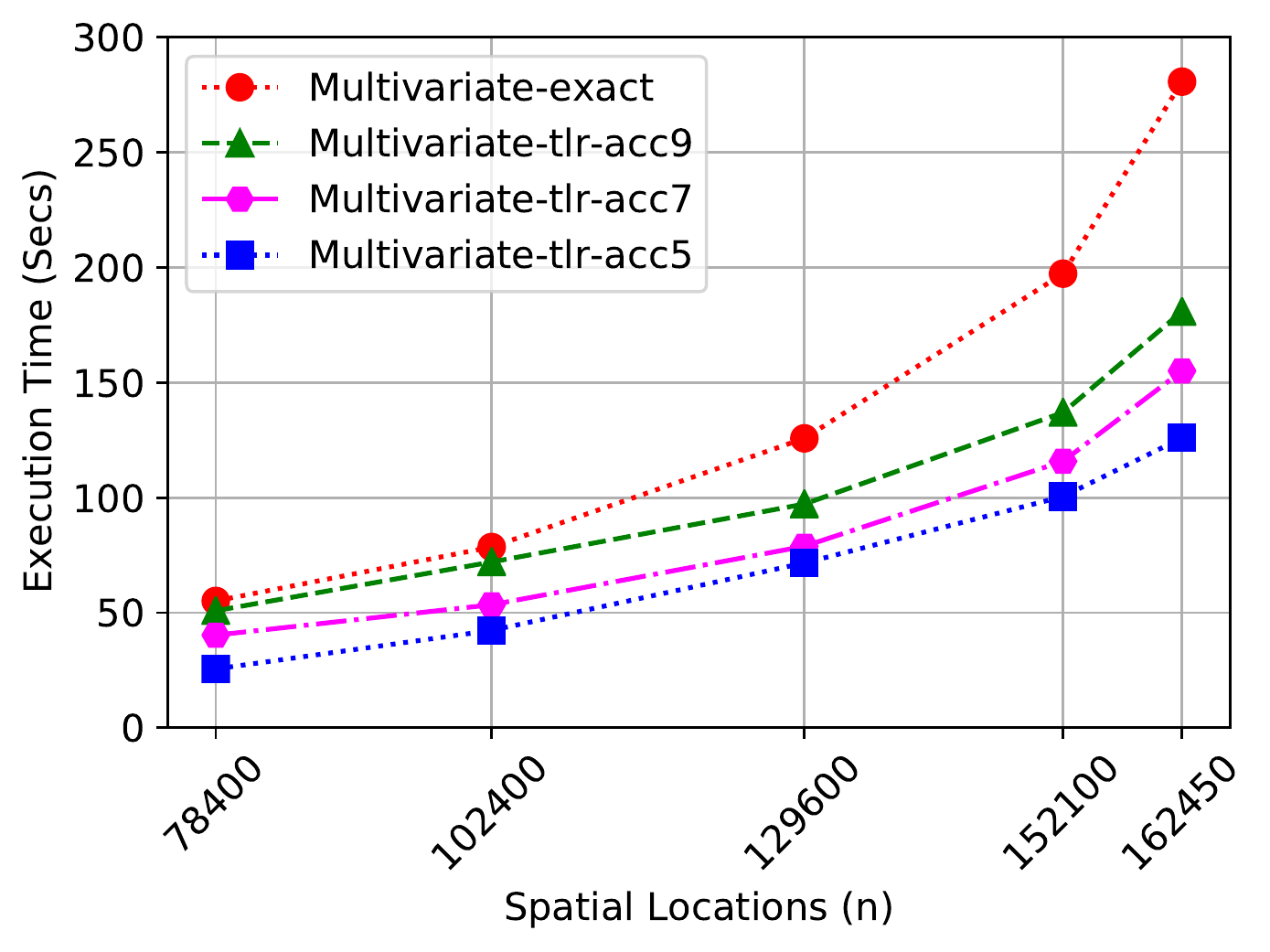}
        \caption{Cray XC40 - 64 nodes}
                \label{fig:64nodes}
    \end{subfigure}%
    ~ 
    \begin{subfigure}[t]{0.23\textwidth}
        \centering
        \includegraphics[height=1.25in]{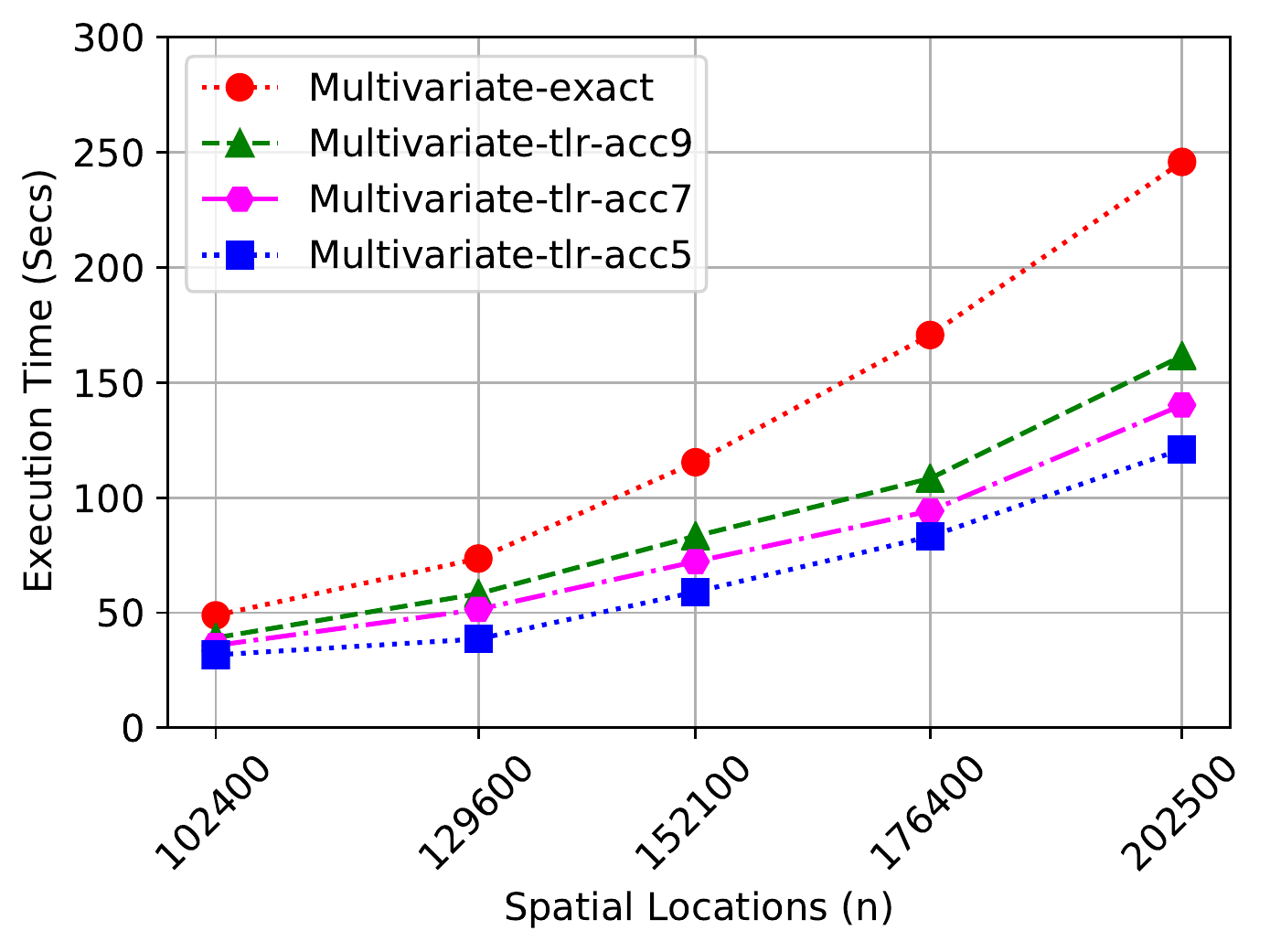}
        \caption{Cray XC40 - 128 nodes}
        \label{fig:128nodes}
    \end{subfigure}
    \caption{One bivariate exact and TLR-based MLE iteration  using different matrix sizes on various shared-memory systems.}
\end{figure}

\begin{figure}[t!]
    \centering
    \begin{subfigure}[t]{0.23\textwidth}
        \centering
        \includegraphics[height=1.25in]{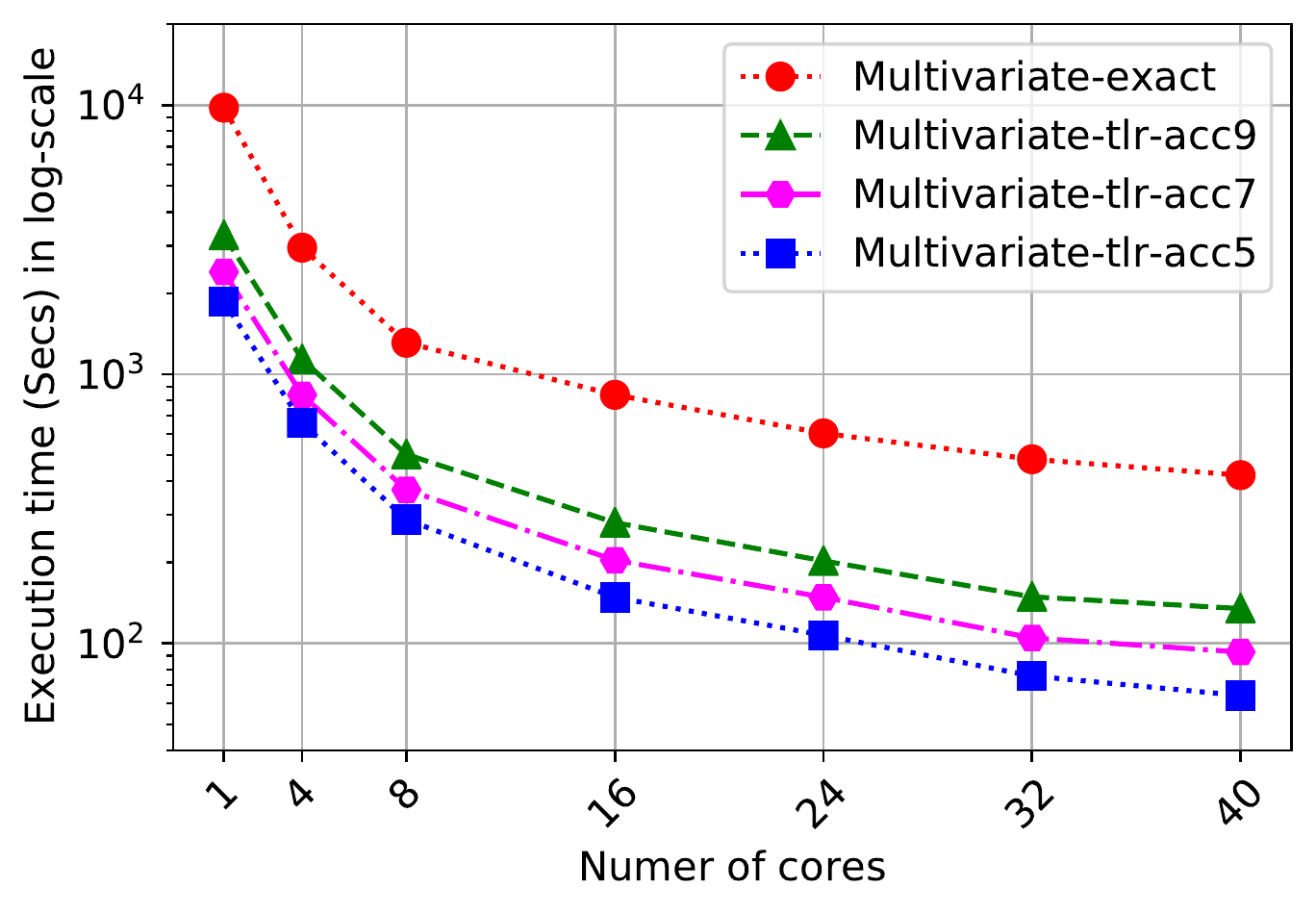}
        \caption{40-core Intel Cascade Lake with different number of cores.}
                \label{fig:cascadelake-scalability}
    \end{subfigure}%
    ~ 
    \begin{subfigure}[t]{0.23\textwidth}
        \centering
        \includegraphics[height=1.25in]{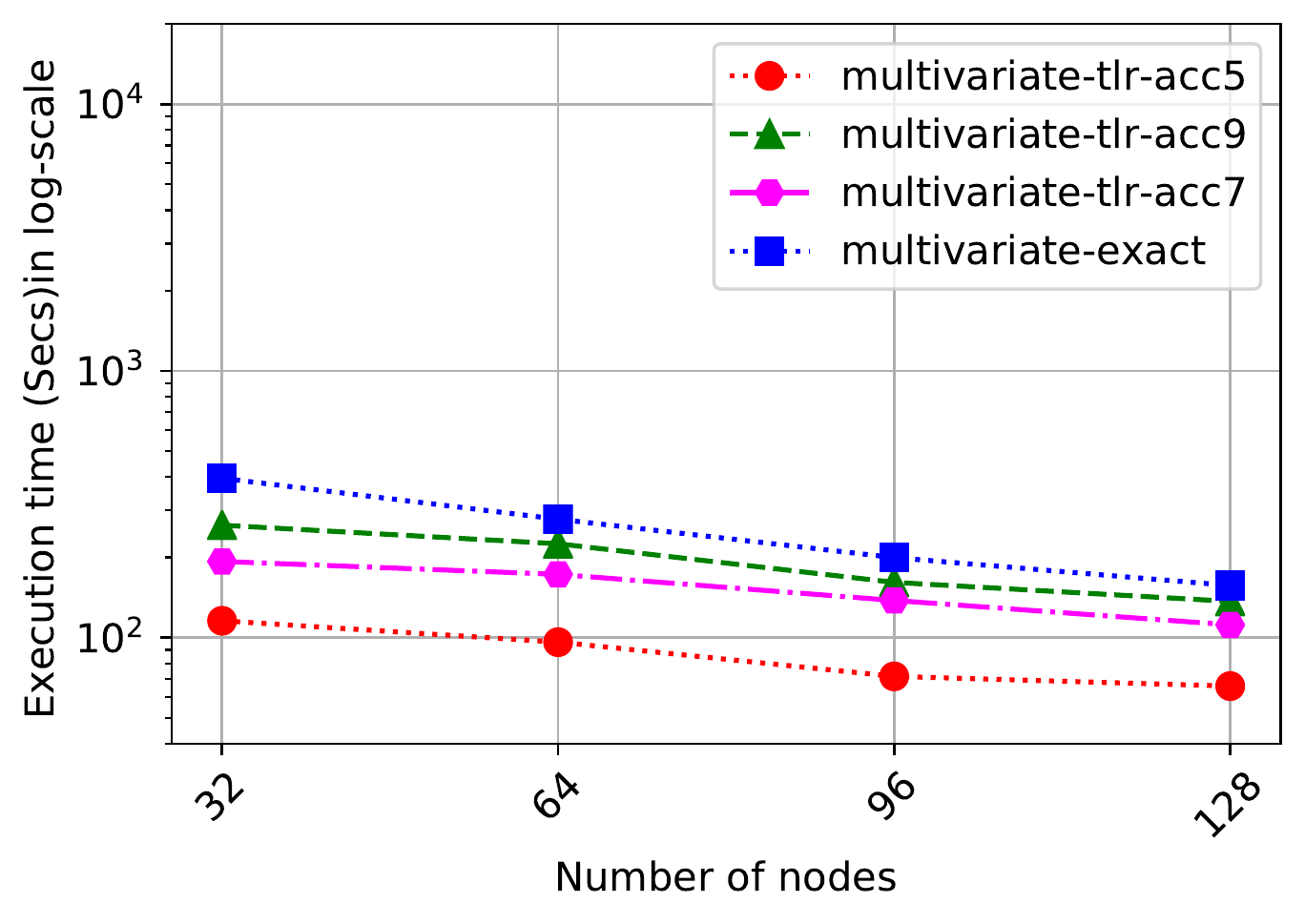}
        \caption{Cray XC40 with different number of nodes.}
        \label{fig:shaheen-scalability}
    \end{subfigure}
    \caption{Strong scalability plot of one bivariate exact and TLR-based MLE iteration.}
     \label{fig:scalability}
\end{figure}

All experiments described in this paper were performed using a variety of shared-memory systems, including a dual-socket 28-core Intel Skylake Intel Xeon Platinum 8,176 CPU running at 2.10 GHz, a dual-socket 20-core Intel Cascade Lake Intel Xeon Gold 6,248 CPU running at 2.50 GHz, a dual-socket 64-core AMD EPYC (Rome) 7702, a dual-socket 20-core Intel Skylake/V100 GPU Intel Xeon Gold 6,148 CPU running at 2.40 GHz, and 32-core ARM ThunderX2 Cavium at 2.10 GHz. For the distributed-memory experiments, we use a Cray XC40 system, with 6,174 16-core Intel dual-socket   Haswell processors running at 2.3 GHz, where each node has 128 GB of DDR4 memory. The KAUST Shaheen-II Cray XC40 system has a total of 197,568 processor cores and 790 TB of aggregate memory.

To obtain timing results, we run each simulation three times on
every single hardware with the same configuration and report
the average. We find runtime variations 
between 0.1\% and 0.5\% on shared-memory systems and
between 1\% and 3\% on distributed-memory system. The latter is slightly
higher since 
the runs are subject to network fluctuations depending on the current
load of the system.

\subsection{TLR-Based Bivariate MLE Performance}

\begin{figure*}[t!]
  
        \scalebox{0.95}{
\vbox{
\begin{minipage}{19cm}
  \centering
    \begin{subfigure}[t]{0.23\textwidth}
        \centering
        \includegraphics[height=1.23in]{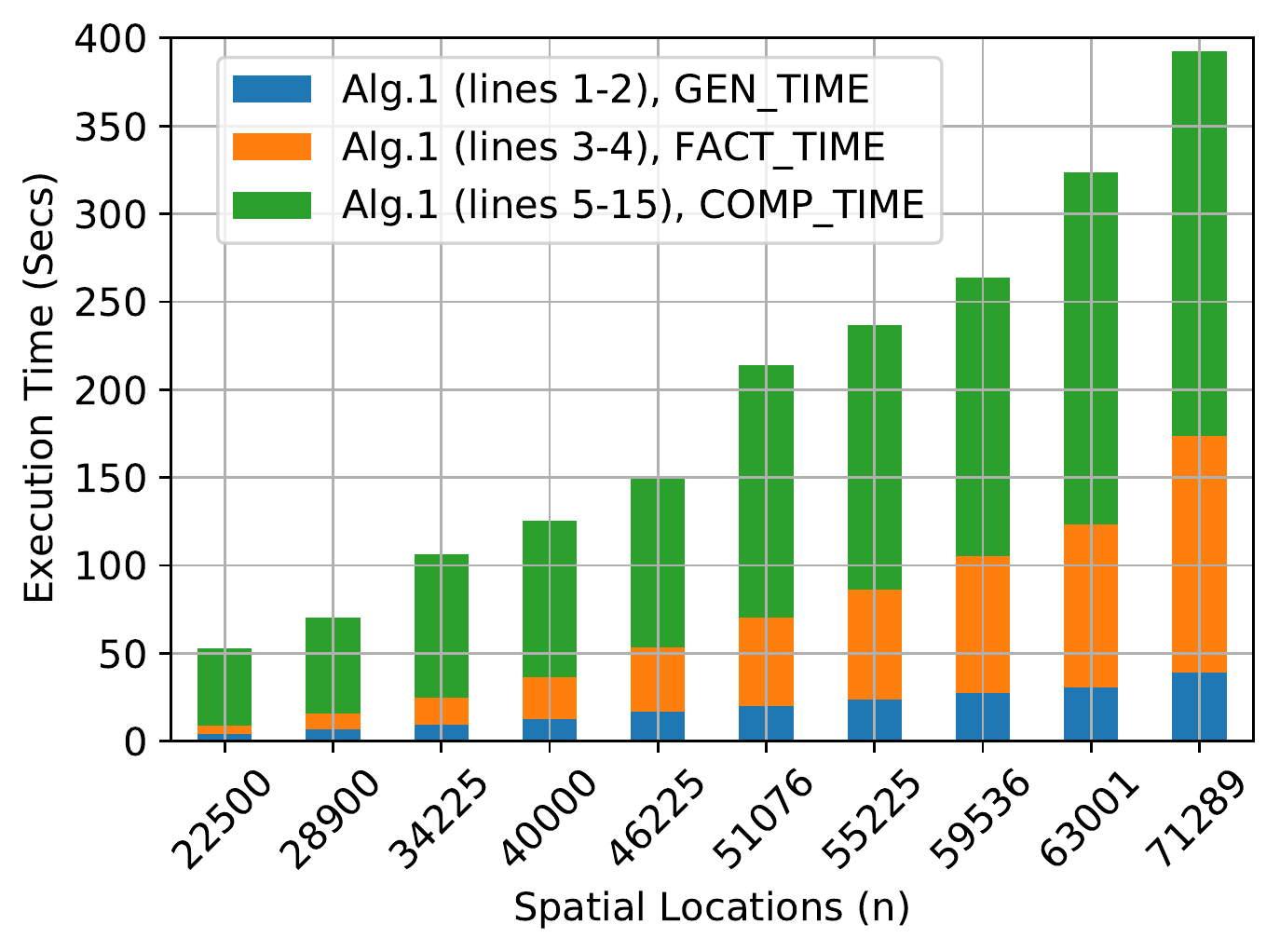}
        \caption{56-core Intel Skylake}
    \end{subfigure}%
    ~ 
    \begin{subfigure}[t]{0.23\textwidth}
        \centering
        \includegraphics[height=1.23in]{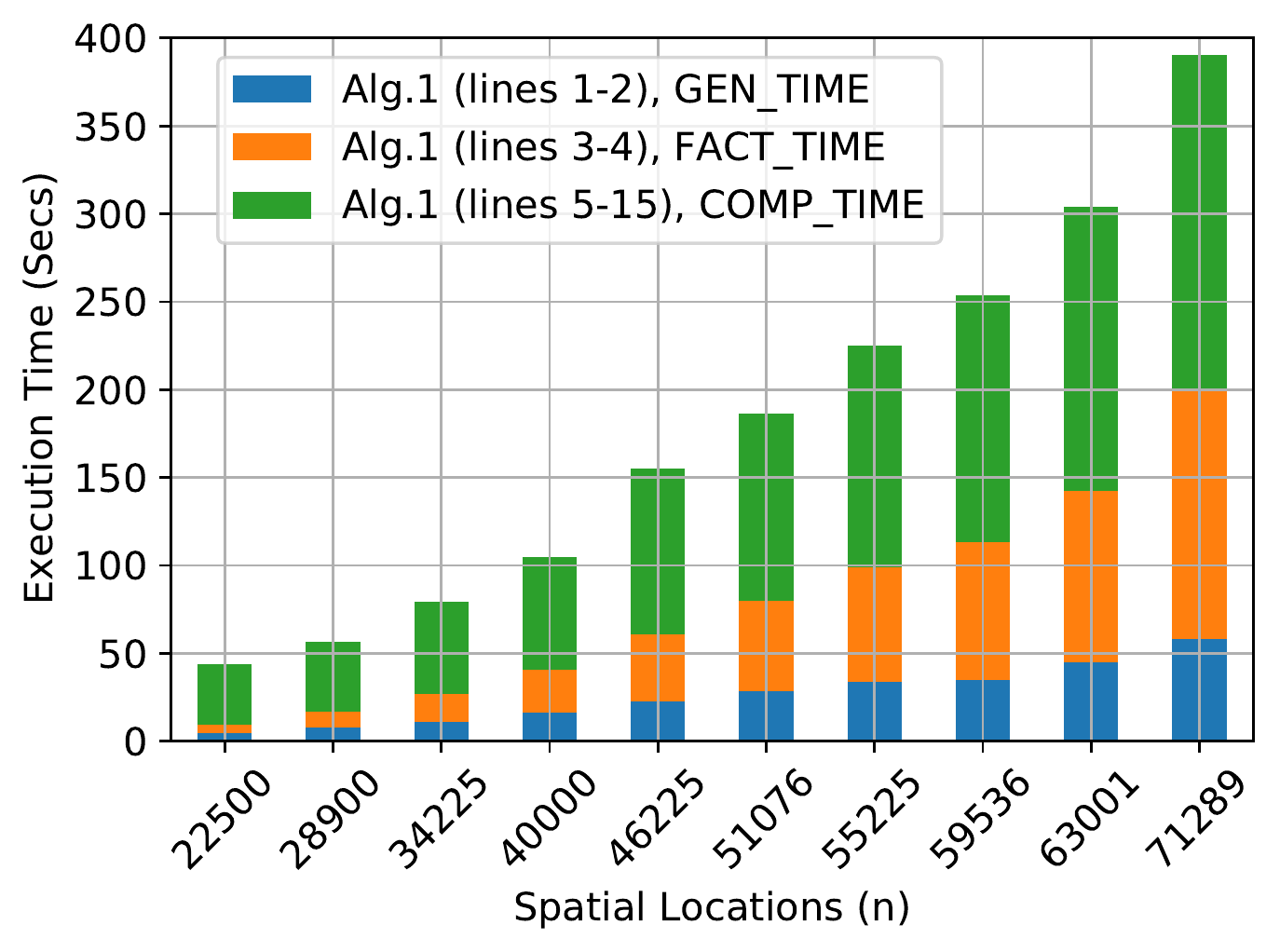}
        \caption{40-core Intel Cascade Lake}
    \end{subfigure}
    ~
        \begin{subfigure}[t]{0.23\textwidth}
        \centering
        \includegraphics[height=1.23in]{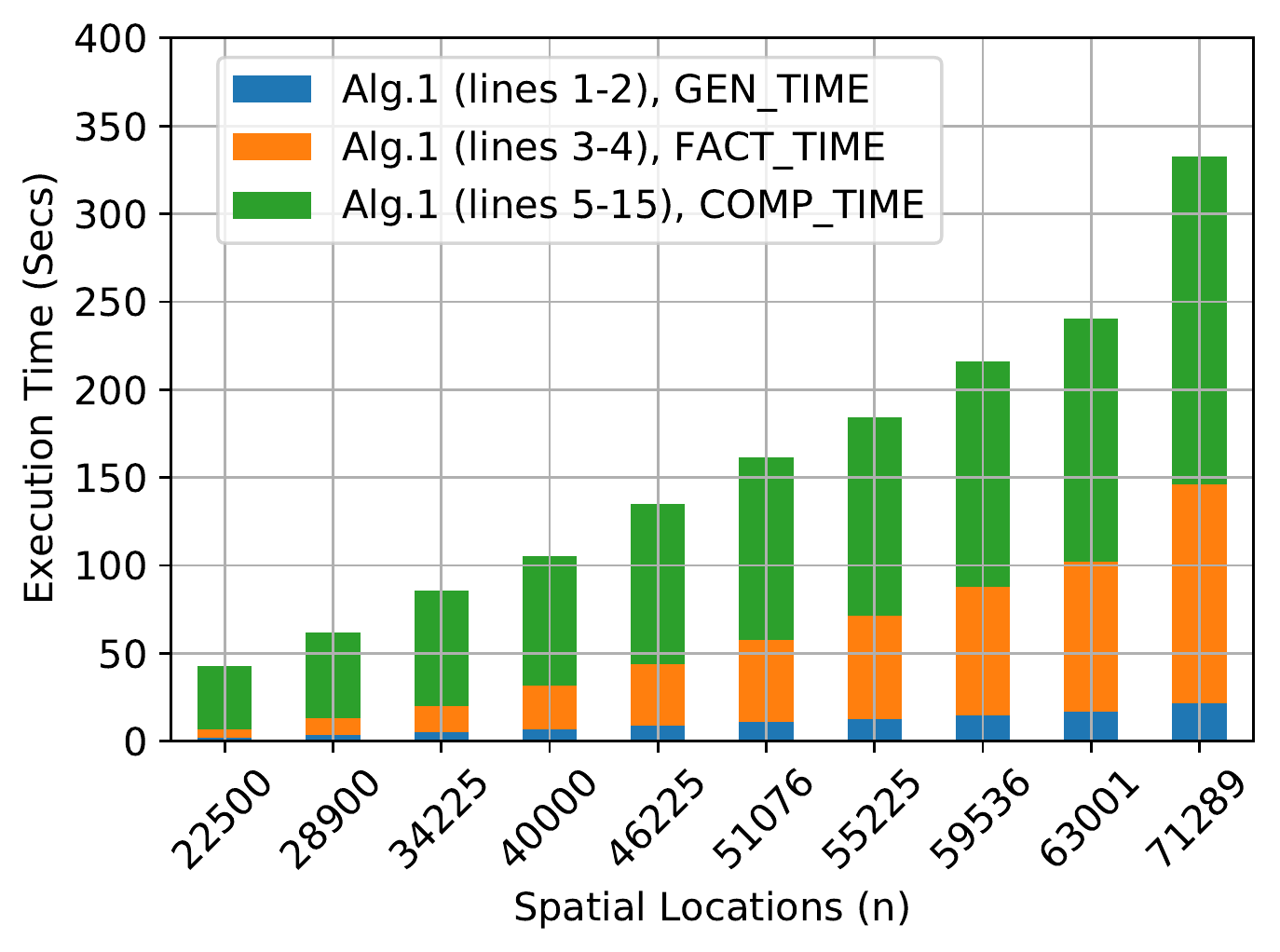}
        \caption{128-core AMD EPYC}
    \end{subfigure}
    ~
            \begin{subfigure}[t]{0.25\textwidth}
        \centering
        \includegraphics[height=1.23in]{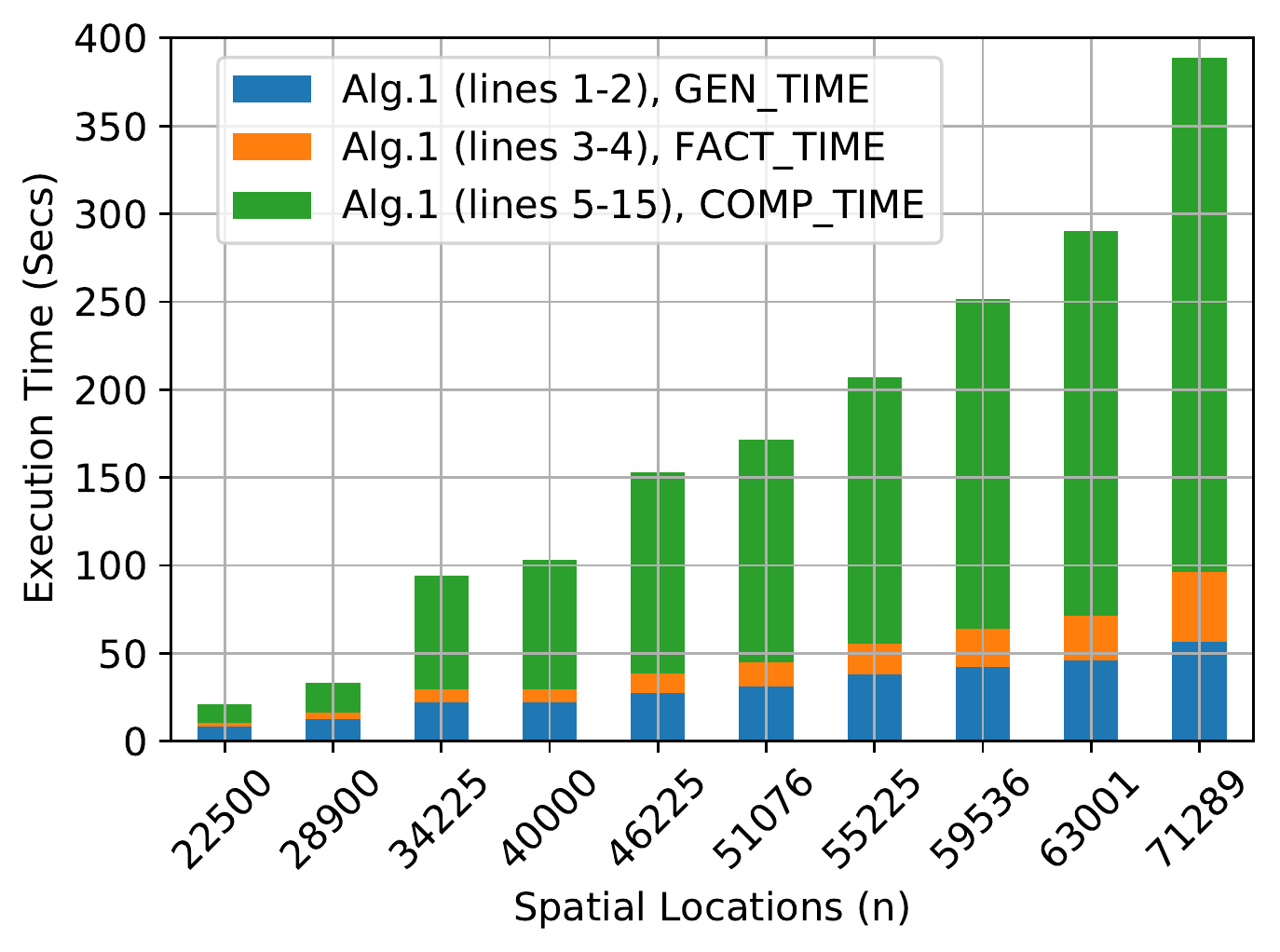}
        \caption{40-core Intel Skylake +V100}
            \label{fig:mloe-mmom-v100}
    \end{subfigure}
    
    \caption{Time breakdown of the univariate  MLOE/MMOM criteria using 100 missing locations on different shared-memory systems.}
    \label{fig:mloe-mmom-performance}
    \end{minipage}
    }
    }
\end{figure*}

\begin{figure*}[t!]
  
        \scalebox{0.95}{
\vbox{
\begin{minipage}{19cm}
  \centering
    \begin{subfigure}[t]{0.23\textwidth}
        \centering
        \includegraphics[height=1.23in]{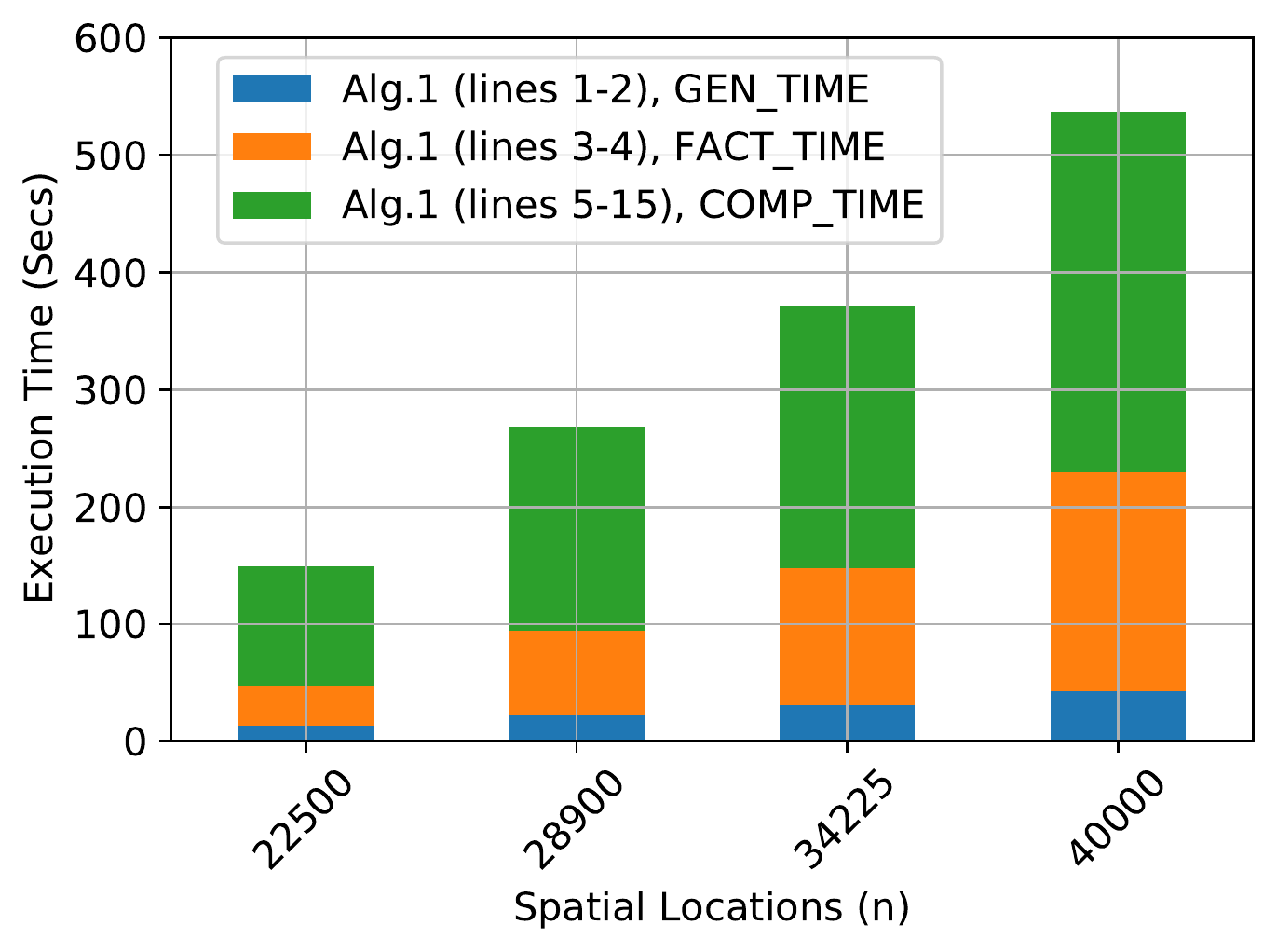}
        \caption{56-core Intel Skylake}
    \end{subfigure}%
    ~ 
    \begin{subfigure}[t]{0.23\textwidth}
        \centering
        \includegraphics[height=1.23in]{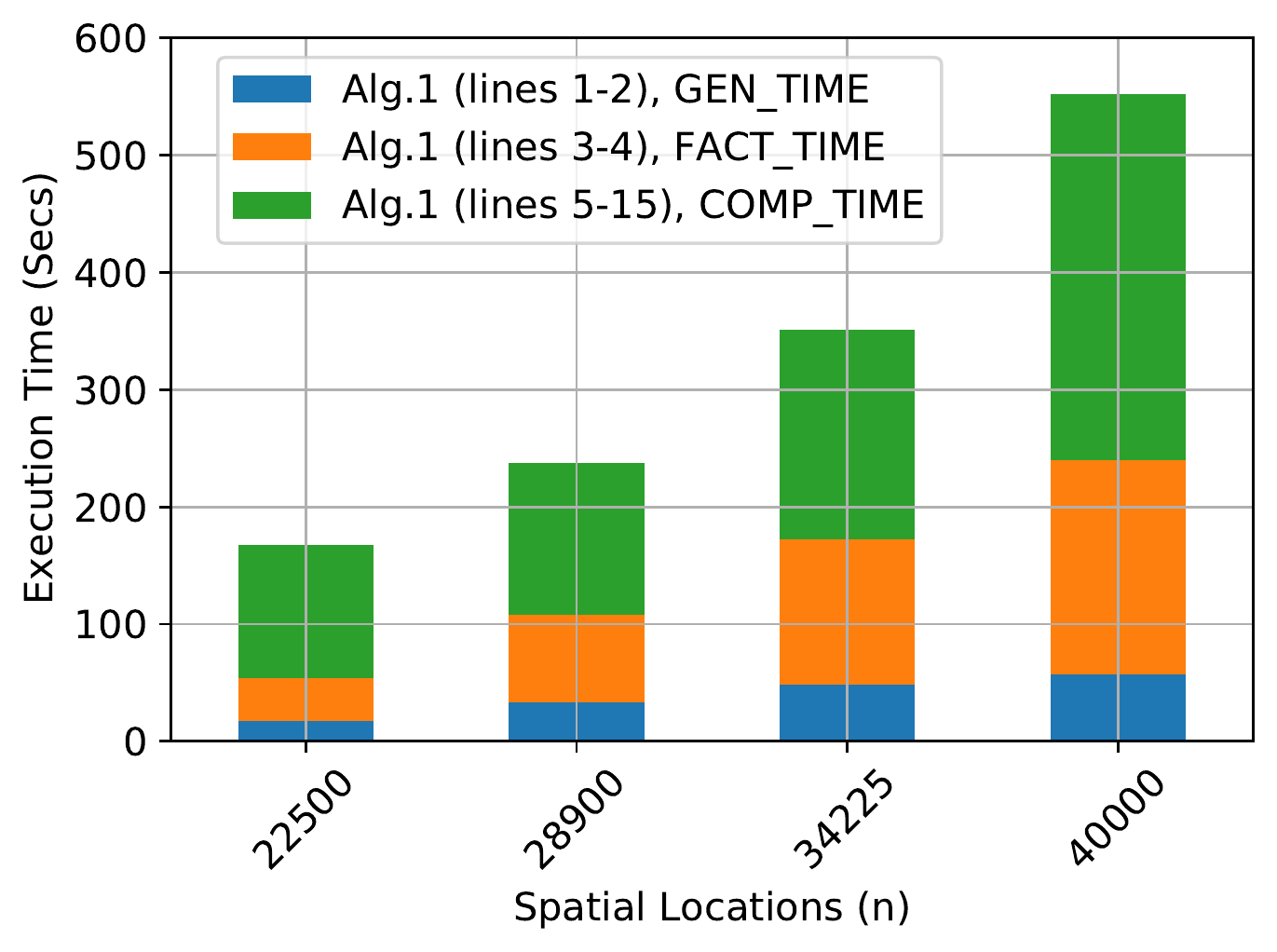}
        \caption{40-core Intel Cascade Lake}
    \end{subfigure}
    ~
        \begin{subfigure}[t]{0.23\textwidth}
        \centering
        \includegraphics[height=1.23in]{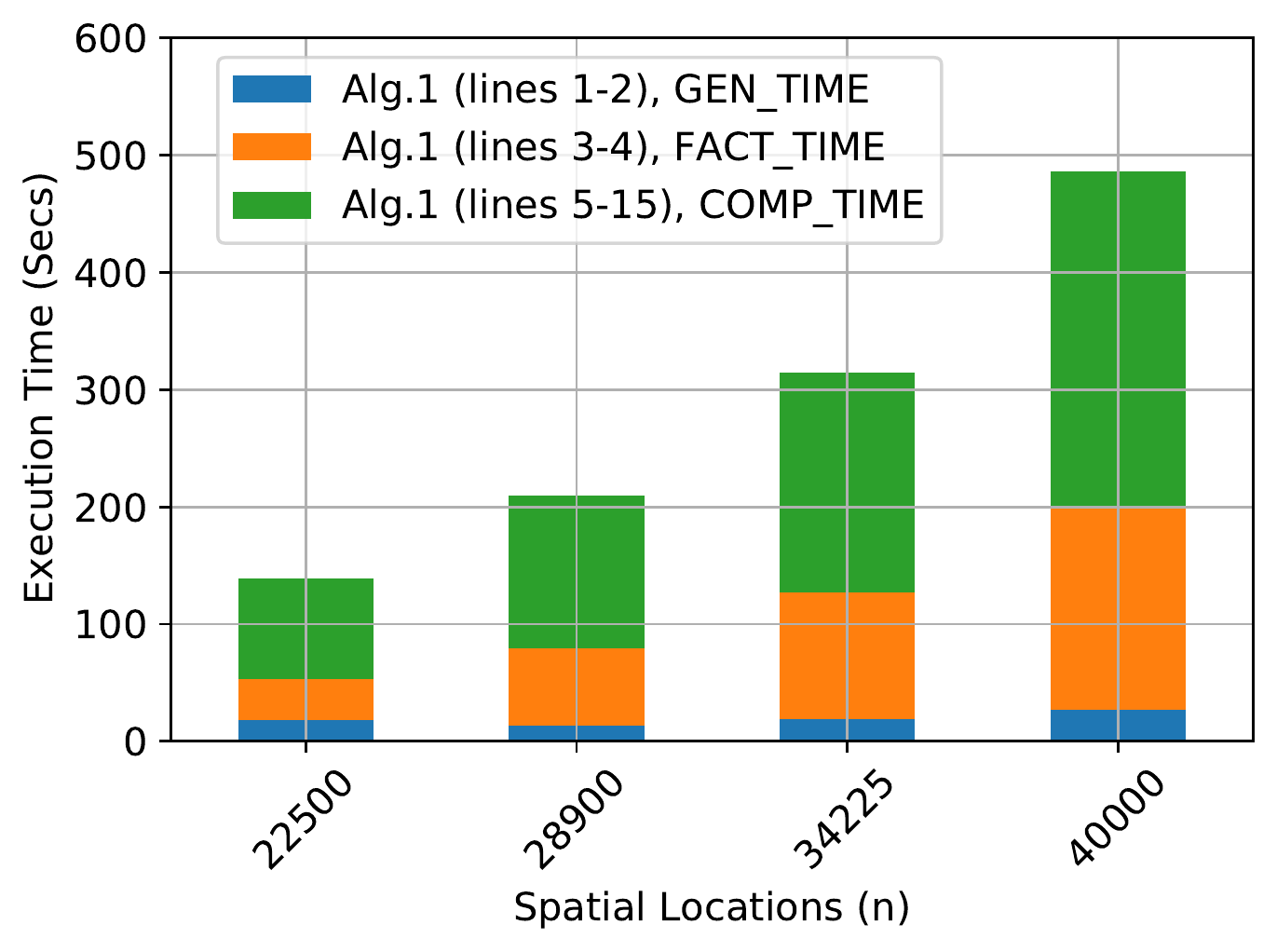}
        \caption{128-core AMD EPYC}
    \end{subfigure}
    ~
            \begin{subfigure}[t]{0.25\textwidth}
        \centering
        \includegraphics[height=1.23in]{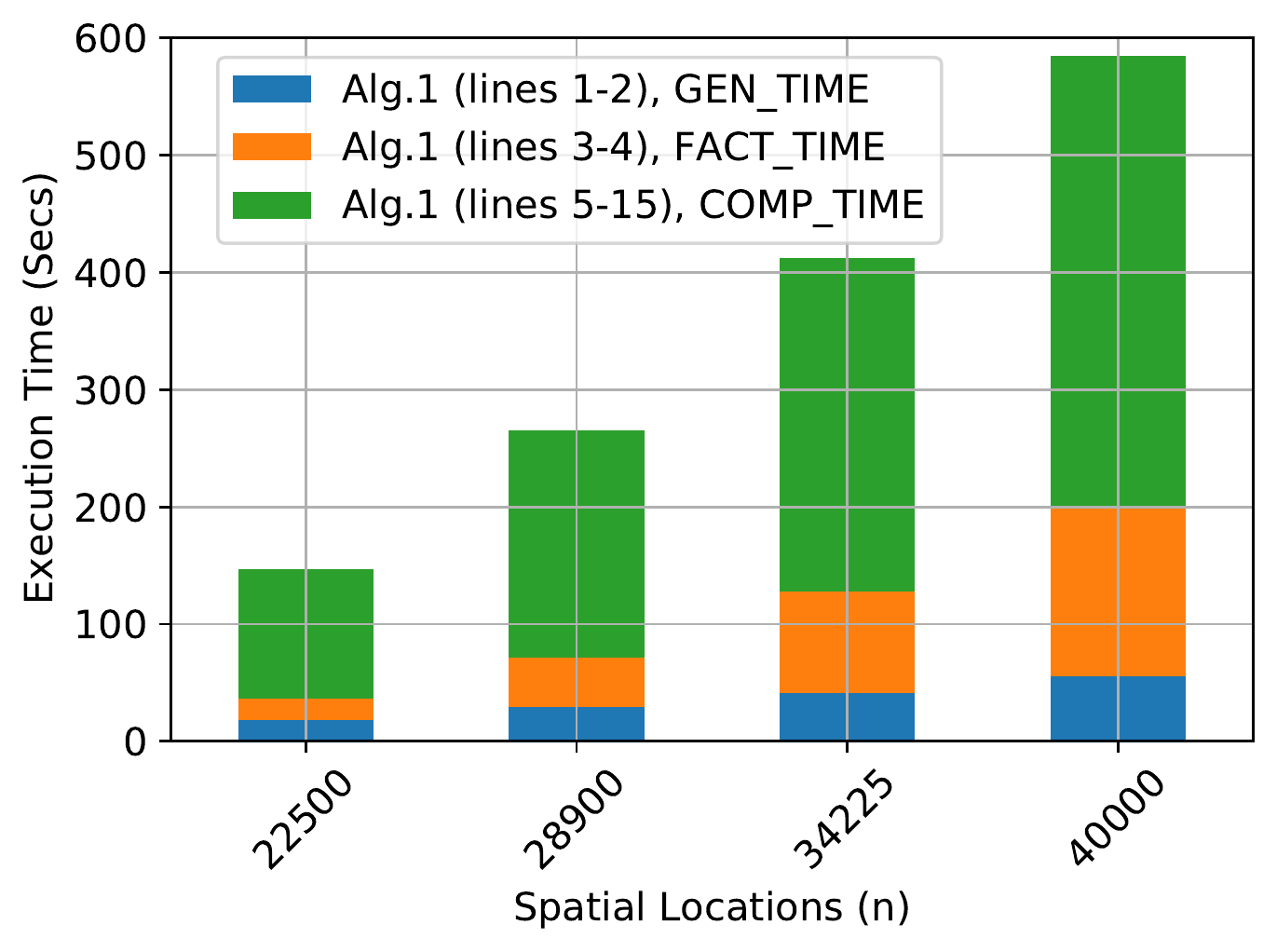}
        \caption{40-core Intel Skylake +V100}
    \end{subfigure}
    
    \caption{Time breakdown of the bivariate  MLOE/MMOM criteria using 100 missing locations on different shared-memory systems.}
    \label{fig:mloe-mmom-performance2}
    \end{minipage}
    }
    }
\end{figure*}

In this section, we evaluate the TLR-based multivariate MLE performance and compare it to the performance of the exact MLE.
All the experiments show one iteration of the MLE
optimization since all the iterations have the same complexity for both exact and TLR-based computation. Fig.~\ref{fig:tlr} shows the
TLR performance on different shared-hardware architectures. The execution
time is shown in the Y-axis (identical for all hardware) while the number of spatial locations is shown
in the X-axis. We use TLR5 as the benchmark of the speedup gained by the TLR computation.

With an Intel 56-core Skylake system, TLR5 can achieve on average 4X
speedup compared to exact MLE, while on an Intel 40-core Cascade Lake
system, the average speedup can reach 4.3X. With a 128-core AMD EPYC
(Rome) system and a 64-core ThunderX2 ARM system, the average speedup
reaches to 6X and 5.5X, respectively. All figures reveal more gains
from the TLR-based approximation with larger problem sizes. Moreover,
with a larger number of cores, the average speedup factor achieved increases.

To compare the gained speedup from each hardware architecture in Fig.~\ref{fig:tlr},
we use the same number of locations $n = 63,001$ as a reference. 
The execution time of one full bivariate TLR-based MLE iteration (TLR5) is 
61.38, 65.65, 35.75, and 113.21 seconds on Intel Skylake, Intel Cascade lake, 
AMD EPYC (Rome), and ThunderX2 ARM chips, respectively. From the speedup perspective,
AMD EPYC chip achieves the best performance compared to
the other systems. It obtains  1.7X, 1.8X, and 3.17X speedup compared to  
Intel Skylake, Intel Cascade lake,  and ThunderX2 ARM systems.

The target hardware systems have different number of cores which
makes comparing them more difficult. 
Given that the implementation is memory-bound, 
we rely on the sustained bandwidth on
each system to give more insights into the
obtained performance. Intel Skylake, Intel Cascade lake, 
AMD EPYC (Rome), and ThunderX2 ARM have a 
sustained bandwidth (measured by STREAM benchmark \cite{STREAM})
178,  140, 300, 236 GB/s. Based on these memory bandwidth values,
we can expect that AMD EPYC (Rome) satisfies 300/178 = 1.68X speedup
compared to Intel Skylake, 300/140 = 2.14X compared to Intel Cascade Lake, and
1.3X compared to ThunderX2 ARM. The calculations show close
values to the obtained speedup except for the ThunderX2 ARM chip.
The speedup discrepancy shows that  
more parallelism needs to
be exposed to the ThunderX2 ARM chip to
take more advantage of its 64 cores.

On the Cray XC40 distributed-memory system, TLR achieves lower speedup compared to the exact but still outperforms it with different problem size and number of nodes. 
Tuning the tile size ($nb$) is challenging on distributed-memory systems and it seems that our baseline
runtime systems, i.e., \starpu, impacts performance with a large number of nodes. Fig.~\ref{fig:64nodes} shows the performance of different TLR accuracy with problem size up to 325K on 64 nodes. The average speedup gained is about 2X.  With 128 nodes, the average speedup gained is about 1.8X as shown by Fig.~\ref{fig:128nodes}.

Fig.~\ref{fig:scalability} shows the strong scalability results using
single node 40-core Intel Cascade Lake system with different number
of cores and Cray XC40 machine using different number of nodes
(up to 128 nodes). 
In Fig.~\ref{fig:cascadelake-scalability},
the Cascade Lake system shows decent parallel speedup as we increase 
the numbers of threads with $n=63,001$. The parallel efficiency on average is 
around $72\%$, (i.e., $T_{min} / ( N * T_N ) * 100\%$, 
where one thread execution time is $T_1$ and $N$ threads execution time
is $T_N$), compared to single thread executions and across different
computation variants, i.e., exact, TLR5, TLR7, and TLR9. In 
Fig.~\ref{fig:shaheen-scalability}, the exact computation achieves
around 66.7X speedup, while the TLR approximation
with different accuracy levels obtains around 51.7X speedup on average with different number of nodes
using $n=168,100$. 
The parallel efficiency of the TLR approximation varies between 60\% and 43\%.
This is lower than the efficiency of the exact computation but expected due to the
memory-bound versus compute-bound regime of executions opposing TLR and exact
computations, respectively.

\subsection{Univariate/Bivariate MLOE/MMOM Criteria Performance}
In this set of experiments, we aim at assessing the performance
of the proposed multivariate MLOE/MMOM criteria algorithm.
We choose the bivariate as an example of a multivariate case
with synthetic datasets generated by our framework.
The experiments were performed on different shared-memory
hardware architectures. We provide the time breakdown of
the assessment operation of Algorithm 1 and split them into three parts: matrices generation
time ($GEN\_TIME$) in lines 1-2, factorization time ($FACT\_TIME$) in lines 3-4, and computation time
($COMP\_TIME$) in lines 5-15.  Figs.~\ref{fig:mloe-mmom-performance} and~\ref{fig:mloe-mmom-performance2} show the 
time for each operation on both parallel univariate and bivariate MLOE/MMOM implementations.

As shown in both figures, the $COMP\_TIME$ is the most time-consuming
part of Algorithm 1 as it requires an iterative execution loop equal
to the number of missing locations, i.e., 100 in this set of experiments.
However, with a larger matrix size, the $FACT\_TIME$ takes more time for the whole
operation.  One striking observation from Fig.~\ref{fig:mloe-mmom-v100} is that using a system with V100 GPU speeds up the $FACT\_TIME$ compared to $COMP\_TIME$. This is because the computation part involves several 
matrix-vector operations (i.e., Level-2 BLAS) that cannot exploit the computational power of the enclosed GPU.

\subsection{TLR-Based Bivariate MLE Accuracy Assessment}
 \label{sec:accuracy_assessment}
Here, we assess the accuracy of the proposed TLR-based bivariate MLE using synthetic datasets.
\subsubsection{Synthetic Datasets}

\begin{figure}[h]
    \centering
    \includegraphics[width=0.5\textwidth]{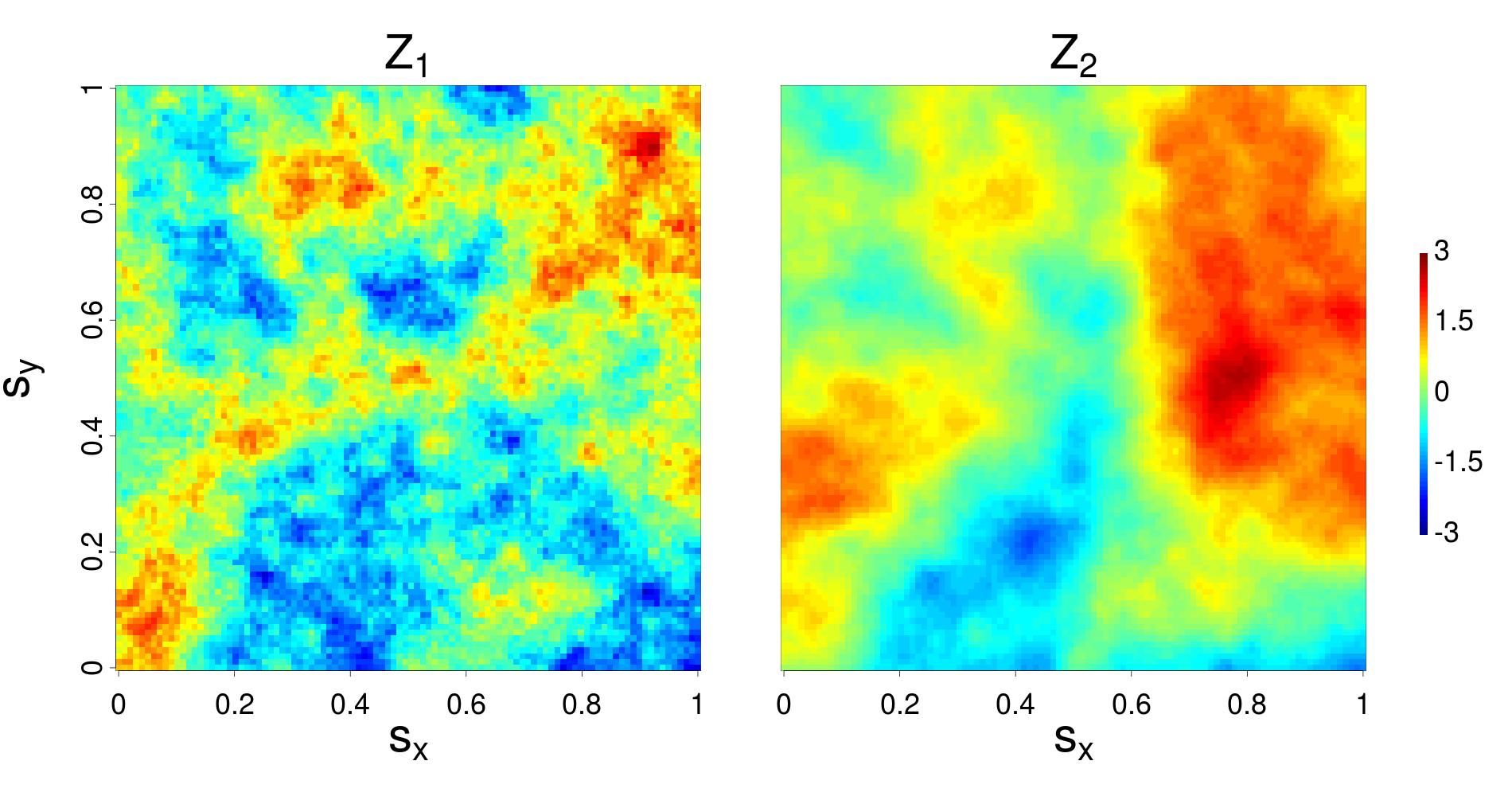}
    \caption{Spatial images of the simulated bivariate realizations from the parsimonious Mat\'{e}rn cross-covariance model on a $158 \times 158$ regular grid on a unit square. Here $\sigma_{11}^2 = \sigma_{22}^2 = 1$, $a = 0.2$, $\nu_{11} = 0.5$, $\nu_{22} = 1$, and $\beta = 0.5$.}
    \label{fig:simulated_Z}
\end{figure}

We perform large-scale simulations from the parsimonious Mat\'e{r}n cross-covariance function. Fig.~\ref{fig:simulated_Z} shows a bivariate random field simulated from Equation~(\ref{eqn:multivariate_matern}) at $n = 24,964$ locations using our synthetic data generator with the following configuration:
\begin{itemize}
    \item $\beta = 0.5$, i.e., $Z_1$ and $Z_2$ are positively correlated. This parameter controls how correlated $Z_1$ and $Z_2$ at any location. The effect of this parameter is visually detectable since wherever there are red (blue) spots in $Z_1$, red (blue) spots in $Z_2$ tend to also be seen.
    \item $\nu_{11} = 0.5$ and $\nu_{22} = 1$, i.e., $Z_2$ is smoother than $Z_1$. The smoothness parameters show through observing that the values of $Z_2$ changes more slowly than the values of $Z_1$ from one pixel or location to another.
    \item $a = 0.2$. This parameter affects $Z_1$ and $Z_2$ in different ways. For $Z_1$, this value of the scale parameter suggests that the marginal covariance of $Z_1$ drops to 0.05 when the locations are 0.6 units apart. For $Z_2$, it takes 0.8 units separation for its marginal covariance to drop to 0.05. Visually, this parameter dictates the sizes of the red and blue spots. The larger the $a$ becomes, the bigger the sizes of the spots are. 
\end{itemize}

\begin{figure}[t!]
    \centering
    \includegraphics[width=0.5\textwidth]{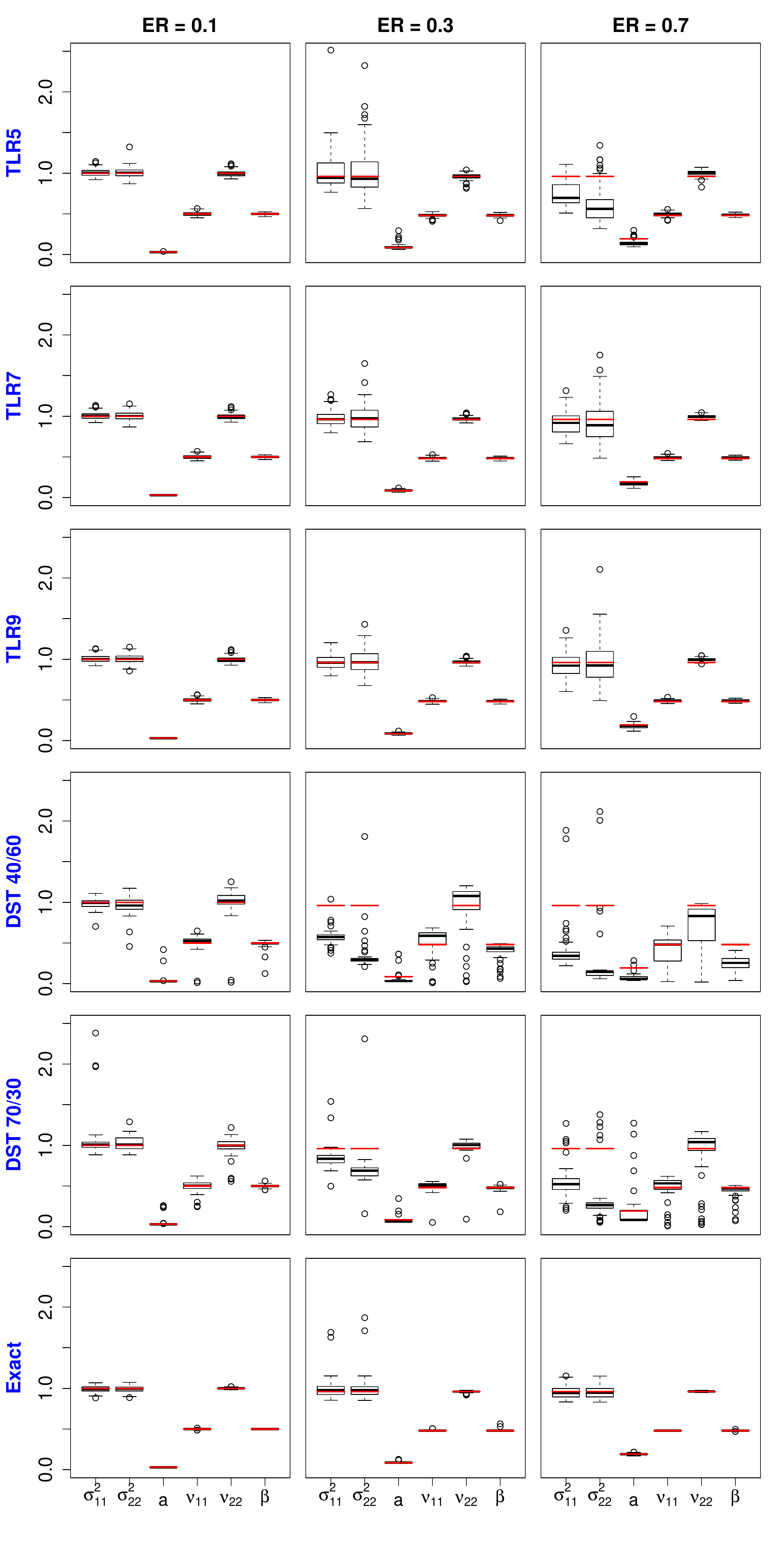}
    \caption{Boxplots of parameter estimates under the exact, TLR, and DST implementations. The true parameters are highlighted in red. ER refers to \textit{effective range} or the distance at which the marginal covariance drops to approximately 0.05. Here ER $= \left\{0.1, 0.3, 0.7\right\}$ corresponds to $a = \left\{0.03, 0.09, 0.2\right\}$.}
    \label{fig:boxplots1}
\end{figure}

\begin{figure}[t!]
    \centering
    \includegraphics[width=0.5\textwidth]{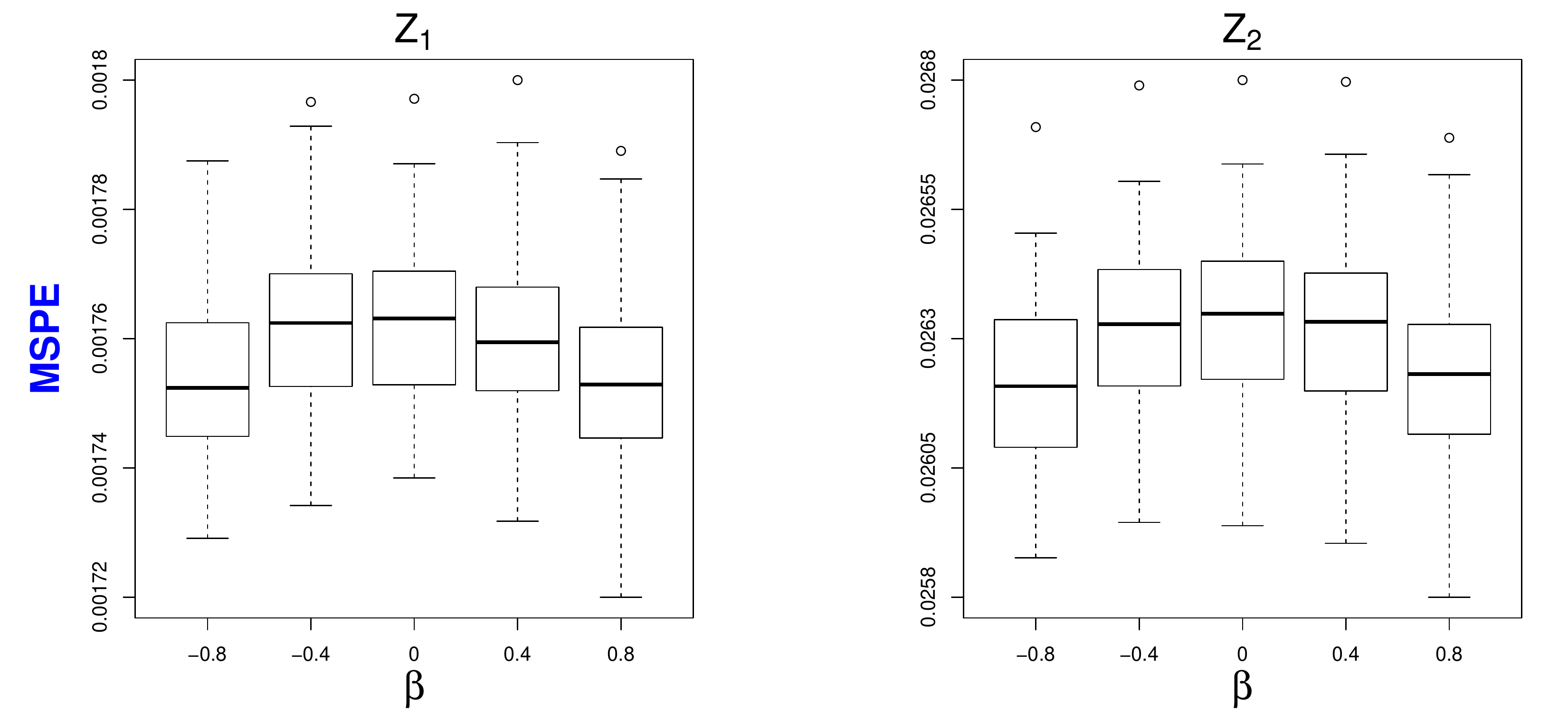}
    \caption{Prediction error (MSPE) at different values of $\beta$ with $\sigma_{11}^2 = \sigma_{22}^2 = 1, \nu_{11} = 0.5, \nu_{22} = 1,$ and $a = 0.09$.}
    \label{fig:boxplots3}
\end{figure}

\begin{figure}[t!]
    \centering
    \includegraphics[width=0.5\textwidth]{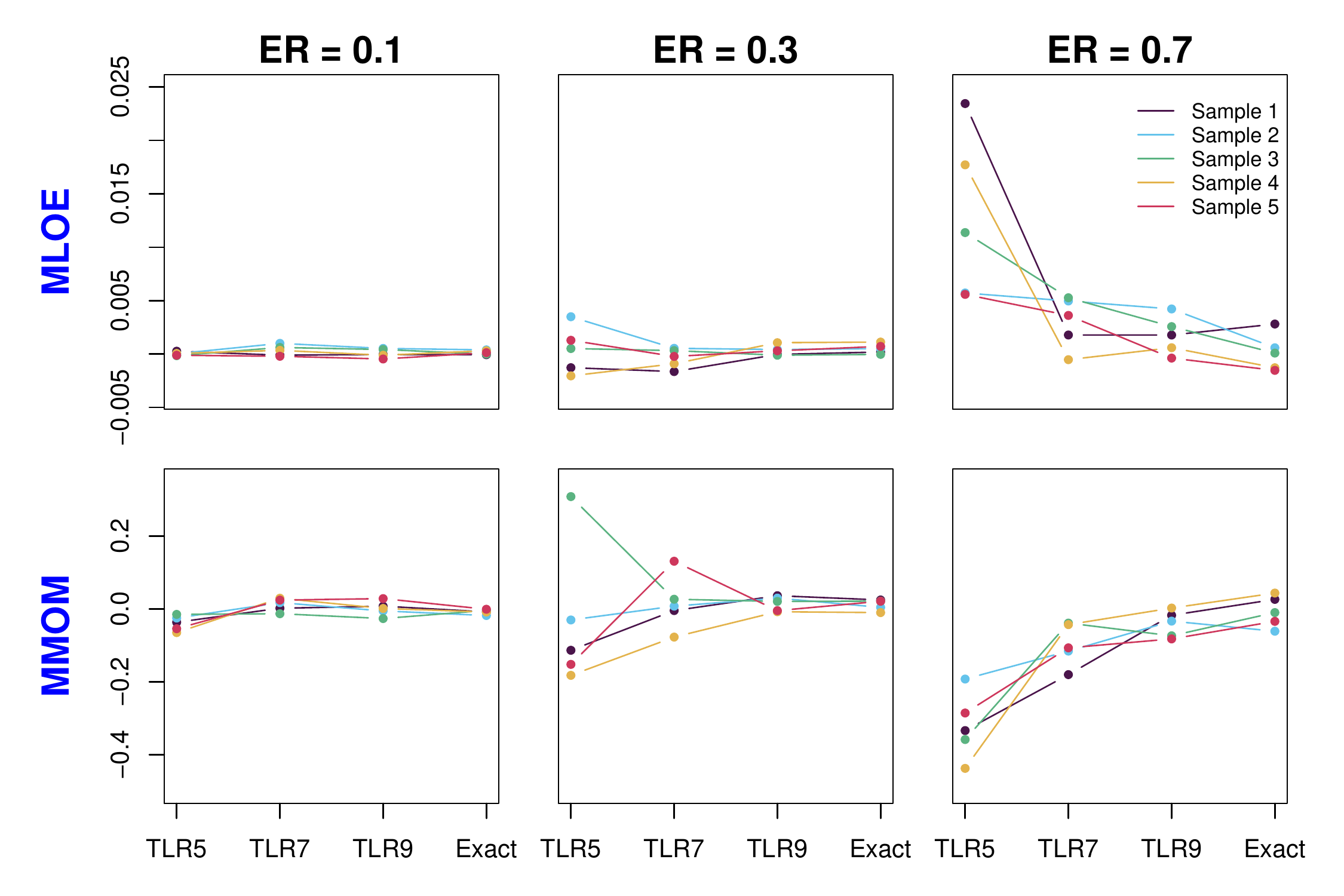}
    \caption{Multivariate MLOE/MMOM values under exact and the different TLR models at various effective ranges.}
    \label{fig:boxplots2}
\end{figure}

To validate the accuracy of the TLR approximation, we simulate 50 different bivariate Gaussian random fields of the same configuration as the example in Fig.~\ref{fig:simulated_Z} and perform three kinds of experiments:
\begin{itemize}
\item Experiment 1: We show the merits of bivariate spatial modeling in the exact computations by varying the degree of colocated dependence between $Z_1$ and $Z_2$, controlled by $\rho$ through $\beta$, and examining whether there is a gain in prediction when using the parsimonious bivariate Mat\'e{r}n for different values of $\beta$, while fixing the other parameters as follows. $\sigma_{11}^2 = \sigma_{22}^2 = 1, \nu_{11} = 0.5, \nu_{22} = 1$, and $a = 0.09$.
\item Experiment 2: We examine the quality of parameter estimates for the exact and TLR under different accuracies (TLR5, TLR7, and TLR9), using the available data in $n = 22,464$ locations (chosen randomly) and reserving the remaining $n_{\text{pred}} = 2,500$ as prediction locations. Furthermore, we contrast these with the estimation results to another approximation technique aforementioned, the Diagonal Super Tile (DST). Fixing $\nu_{11} = 0.5$ and $\nu_{22} = 1$, we vary the value of the remaining parameter responsible for spatial dependence over long distances, i.e., the range parameter, $a$. Different values of $a$ were chosen to represent weak $(a = 0.03)$, moderate $(a = 0.09)$, and strong  $(a = 0.20)$ spatial dependencies.
    \item Experiment 3: Predictions are made at the $2,500$ prediction locations and the errors produced by the exact and the approximation models are assessed using the newly proposed multivariate MLOE/MMOM.
   
\end{itemize}

Fig.~\ref{fig:boxplots3} summarizes the results of Experiment 1. The figure shows that the higher the value of the parameter responsible for the colocated dependence ($\beta$), the lower the prediction error becomes. In bivariate datasets, the inclusion of a second variable effectively increases the number of samples available for any of the two variables when the colocated dependence between them is high (positive or negative). The more correlated $Z_1$ and $Z_2$ are, through $\beta$, the more information we get about $Z_1$ from $Z_2$, and vice versa. This echoes the theoretical results in \cite{zhang2015doesn}, where it was shown that the colocated correlation parameter dictates the improvement introduced by cokriging (multivariate prediction) over kriging (univariate prediction). Similar conclusions were derived using other values of $a$, the parameter controlling the long range spatial dependence. This suggests that bivariate or multivariate ($p>2$) modeling should be pursued regardless because while the gain in using a bivariate model is not so pronounced when the colocated dependence is not so high in the positive or negative direction, there is significant error reduction when the variables turn out to be highly correlated. 

Fig.~\ref{fig:boxplots1}, plots the accuracy of our estimation procedure under the different TLR and exact implementations at different strengths of spatial dependence controlled by the range parameter, $a$. It also includes the results of the parameter estimation under the two different sizes of the DST.  DST 40/60 means that 40\% of the tiles from the diagonal are kept and the remaining 60\% are annihilated. Similarly, DST 70/30 denotes that 70\% of the tiles from the diagonal are kept and the remaining 30\% are annihilated. When the spatial dependence is weak ($a = 0.03$), the boxplots across the different TLRs and the exact are identical, i.e., the medians and the standard deviations of the parameter estimates are almost equal. While the medians of the parameter estimates under DST 40/60 and DST 70/30 are close to the true parameter values, the estimates have more variability. As the dependence in space increase, i.e., $a = 0.09$ (moderate) and $a = 0.2$ (strong), TLR5 is insufficient and it obtains parameter estimates that are very far from the true value. While increasing the accuracy to TLR7 solves the problem for moderate spatial dependence, this level of accuracy is still inadequate for simulations with strong spatial dependence as the medians of the parameter estimates still do not coincide with the true parameter values. TLR9 remedies this problem. DST 40/60 and DST 70/30 give estimates that are far from the true values and they perform worse as the strength of spatial dependence increase. This is expected as the DST technique throws away significant amount of information in the cross-covariance matrix which is vital when the dependence in space is strong. All in all, while there are significantly more parameters to estimate in the parsimonious bivariate Mat\'{e}rn model, our estimation procedure can satisfactorily recover all of them. Furthermore, TLR approximation outperforms  another approximation technique, i.e., the DST, and remains competitive with the exact model in terms of parameter estimation accuracy when using a higher accuracy level whenever there is stronger spatial dependence.

Using the parameter estimates in Experiment 2, we predict at the $2,500$ unsampled locations and measure the multivariate MLOE/MMOM for all the TLR approximation models and the exact using Algorithm 1. Fig.~\ref{fig:boxplots2} charts the behavior of the multivariate MLOE/MMOM of 5 randomly chosen sample bivariate random fields as we increase the accuracy levels. Commensurate to the findings in Experiment 2, bivariate random fields with higher spatial dependence necessitate TLR approximations with higher accuracy levels in order to remain competitive with the exact model. 

\subsubsection{Real Data Application}

We obtained datasets with a horizontal spatial resolution of 5 km from a Weather Research and Forecasting (WRF) model simulation on the $[43^{\circ} E, 65^{\circ}E] \times [5^{\circ}S, 24^{\circ}N]$ region of the earth \cite{yip2018statistical}. We restricted the dataset to the Arabian Sea to ensure that the measurements exhibit spatial isotropy, i.e., the cross-covariance depends only on the distance between any two locations and not on the locations themselves. Often, this isotropy assumption holds when the locations are situated in areas with similar characteristics. As the locations are all on the ocean, this behavior can be expected. The resulting subset contains $n = 116,100$ locations and the two locations which are located farthest from each other have a great circle distance of $2,681$ km. 

\begin{enumerate}[I.]

\begin{figure}[t!]
    \centering
    \includegraphics[width=0.5\textwidth]{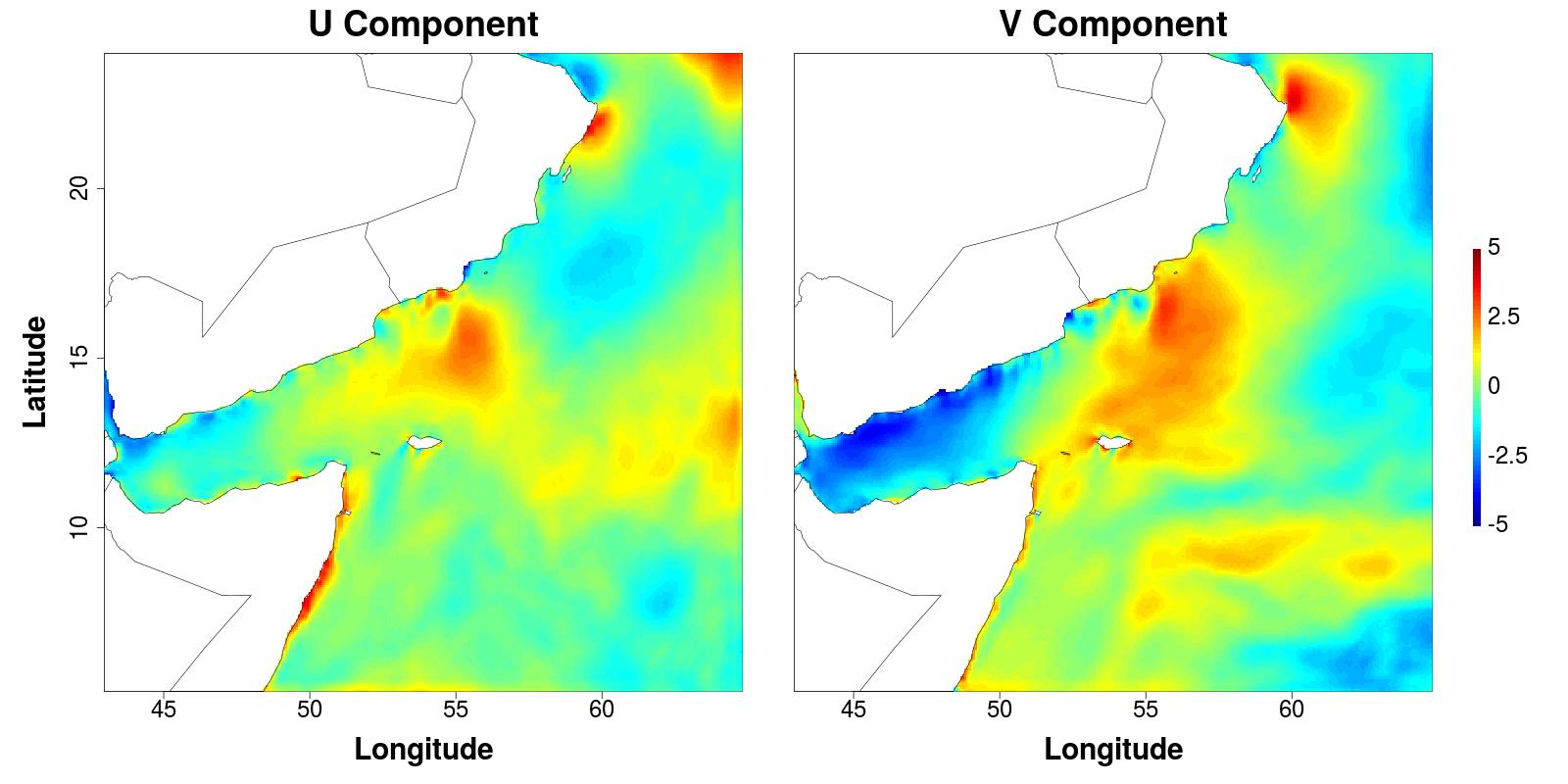}
    \caption{Spatial images of the bivariate dataset on January 1, 2009 (after mean removal) on $116,100$ locations over the Arabian Sea.}
    \label{fig:application_arabian_sea_residuals}
\end{figure}

   \begin{table}[t!]
\caption{A summary of the estimated parameters and the breakdown per variable and average of the prediction error, denoted by MSPE$_i$, $i = 1, 2$, and MSPE$_{\text{avg}}$, respectively, of the bivariate model fitted to the bivariate dataset on January 1, 2009.}
  \label{tab:application_results}
 \caption*{\small Parsimonious Bivariate Mat\'{e}rn} 
    \centering
{\scriptsize
\begin{tabular}{lcccccccccc}
    \toprule    
$\hat{\sigma}_{11}^2$ & $\hat{\sigma}_{22}^2$   & $\hat{a}$ & $\hat{\nu}_{11}$ & $\hat{\nu}_{22}$   & $\hat{\beta}_{12}$   \\
     \midrule
0.718 & 0.710 & 0.161 & 2.283 & 2.033 &  0.192\\
    \bottomrule
  \end{tabular}
 
 \vspace{5mm}
 
  \begin{tabular}{lcccccccccc}
    \toprule    
MSPE$_1$ & MSPE$_2$ & MSPE$_{\text{avg}}$ \\
     \midrule
0.000189  & 0.000261 & 0.000225 \\
    \bottomrule
  \end{tabular}
  }

\end{table}

\begin{figure}[t!]
    \centering
    \includegraphics[width=0.5\textwidth]{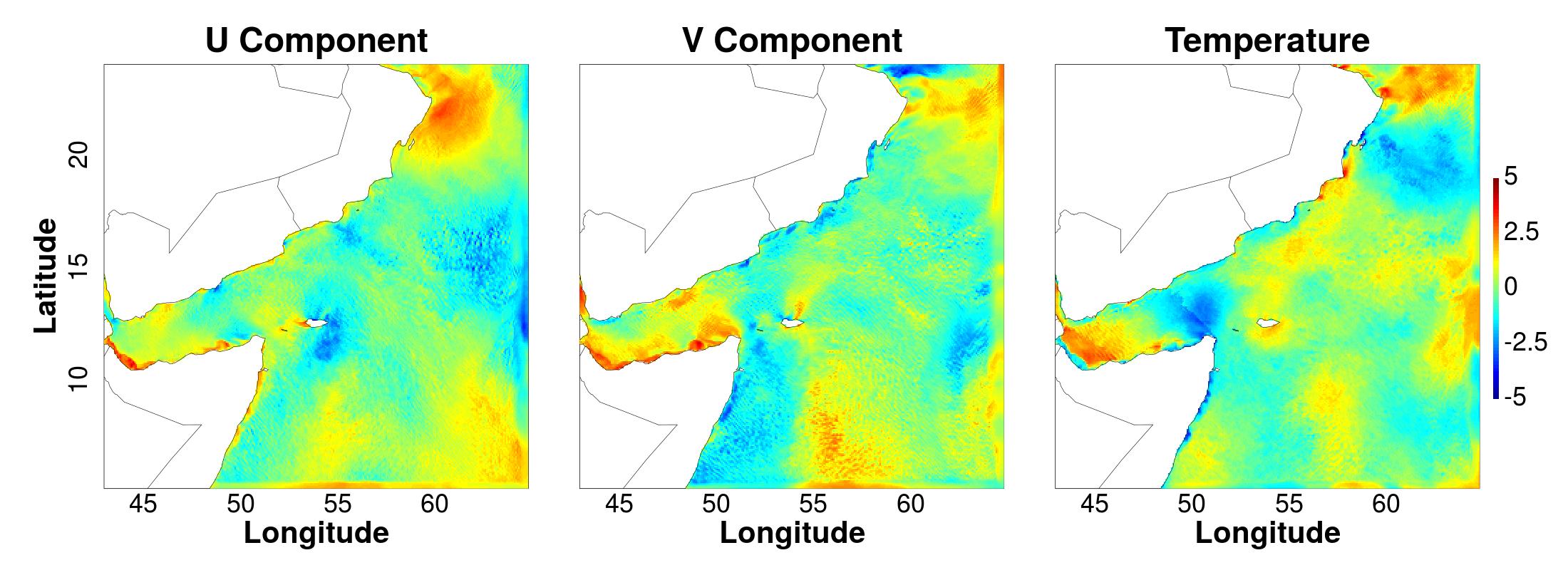}
    \caption{Spatial images of the trivariate dataset on October 1, 2009 (after mean removal) on $116,100$ locations over the Arabian Sea.}
    \label{fig:application_arabian_sea_residuals2}
\end{figure}

 \begin{table}[t!]
\caption{A summary of the estimated parameters and the breakdown per variable and average of the prediction error, denoted by MSPE$_i$, $i = 1, 2, 3$, and MSPE$_{\text{avg}}$, respectively, of the trivariate model fitted to the trivariate dataset on October 1, 2009.}
 \caption*{\small Parsimonious Trivariate Mat\'{e}rn} 
  \label{tab:application_results2}
\centering
{\scriptsize
\begin{tabular}{lccccccccccccccc}
    \toprule    
$\hat{\sigma}_{11}^2$ & $\hat{\sigma}_{22}^2$  & $\hat{\sigma}_{33}^2$ & $\hat{a}$ & $\hat{\nu}_{11}$ & $\hat{\nu}_{22}$  & $\hat{\nu}_{33}$ \\
     \midrule

0.788&0.874&0.301&0.0822&1.689&1.629&
1.234\\
    \bottomrule
  \end{tabular}
  
   \vspace{5mm}
  
  \begin{tabular}{lccccccccccccccc}
    \toprule    

   $\hat{\beta}_{12}$  & $\hat{\beta}_{13}$  & $\hat{\beta}_{23}$  & MSPE$_1$ & MSPE$_2$ & MSPE$_3$ & MSPE$_{\text{avg}}$ \\
     \midrule
0.243&-0.124&-0.059 & 0.009900&0.012248&0.021073&0.014407 \\
    \bottomrule
  \end{tabular}

  }

\end{table}

\item Bivariate Dataset

We fit the parsimonious bivariate Mat\'{e}rn covariance function on measurements obtained on January 1, 2009, consisting of two variables, namely, zonal wind component, $U$ (variable 1), and meridional wind component, $V$ (variable 2), both measured in $m/s$.  In order to satisfy the zero-mean assumption, we remove a spatially varying mean using the longitudes and latitudes as covariates. The resulting values after mean removal are approximately Gaussian and are plotted in Fig.~\ref{fig:application_arabian_sea_residuals}.

The parameter estimates for the parsimonious bivariate Mat\'{e}rn fitted to the dataset at $n = 104,490$ observation 
locations are presented in Table~\ref{tab:application_results}. 
The MSPE values for the predictions done on $n_{\text{pred}} = 11,610$ prediction locations are also shown.
From the results, it can be seen that $U$ and $V$ (variables 1 and 2) are positively correlated since $\hat{\beta}_{12} > 0$. Furthermore, the estimates of the smoothness parameters $\hat{\nu}_{11}$ and $\hat{\nu}_{22}$ suggest that $U$ and $V$ are very smooth random fields, which is certainly the case as shown in Fig.~\ref{fig:application_arabian_sea_residuals}.

\item Trivariate Dataset

We retrieve another dataset from October 1, 2009 consisting of three variables including the $U$ and $V$ wind components (in $m/s$) and temperature (in Kelvin), as variable 3, and fit a trivariate Mat\'{e}rn covariance function. Similarly, we remove a spatially varying mean using the longitudes and latitudes as covariates. The resulting values after mean removal are approximately Gaussian and are plotted in Fig.~\ref{fig:application_arabian_sea_residuals2}.

The parameter estimates for the parsimonious
trivariate Mat\'{e}rn are presented in Table~\ref{tab:application_results2} with the corresponding MSPE values.
It can be seen that $U$ and $V$ (variables 1 and 2) are positively correlated while each of them are negatively correlated to variable 3, the temperature variable. These values obtained for the correlation coefficients can be visually validated by Fig.~\ref{fig:application_arabian_sea_residuals2}. In regions where blue/green spots are observed in the $U$ and $V$ components, red/yellow spots generally occur for the temperature variable, especially in the $[53^{\circ} E, 65^{\circ}E] \times [13^{\circ}S, 17^{\circ}N]$ region. This inverse relationship can also be seen along the coast of Yemen and Somalia. 

\end{enumerate}

Note that the estimates for the smoothness and the range parameters in the bivariate model, 
i.e., $\hat{\nu}_{11}$, $\hat{\nu}_{22}$, and $\hat{a}$, are higher compared to 
its counterpart in the trivariate model. This is expected as the spots are larger and smoother in Fig.~\ref{fig:application_arabian_sea_residuals} than in Fig.~\ref{fig:application_arabian_sea_residuals2}. These foregoing results definitely show that
our proposed implementation was able to successfully fit the models with physically reasonable parameter estimates.

\section{Conclusion} \label{sec:conclusion}
We proposed a high-performance framework for modeling and inference of large spatial datasets based on the multivariate geospatial statistical modeling concept. In the context of climate and weather applications, the framework can operate on the Gaussian log-likelihood function with three (or more) associated variables, for the purpose of estimating a parameter vector, in order to predict missing measurements. 
Although machine learning and deep learning techniques can also be used in prediction, geospatial statistical modeling has better interpretability capabilities regarding the underlying spatial field. 
Both the exact and TLR-based approximation computations of the MLE operations were implemented and evaluated on large-scale experiments. The TLR-based approximation for the MLE outperformed
the fully double-precision exact MLE counterpart up to 10X and 2X on different hardware architectures. Comprehensive qualitative experiments were conducted to assess the accuracy of the TLR-based estimation and prediction. We demonstrated the effectiveness of the approximation technique in achieving high performance, while preserving a convenient
accuracy level. 
Additionally, an algorithm to compute the newly proposed multivariate MLOE/MMOM criteria was devised. This algorithm allows for the assessment of the quality of the MLE operations involving approximated models.

Future research will focus on modeling and prediction of environmental variables which are indexed in space and time. The spatial and temporal coverage of big geospatial data can be exploited to improve insights on an environmental phenomenon. Tackling the space-time problem should bring more challenges related to the problem dimension and prediction accuracy in climate/weather applications.


%

\ifCLASSOPTIONcompsoc
  \section*{Acknowledgements}
\else
  \section*{Acknowledgement}
\fi

The authors would like to thank
NVIDIA Inc., Cray Inc., and Intel Corp., the Cray Center of
Excellence and Intel Parallel Computing Center awarded to
the Extreme Computing Research Center (ECRC) at KAUST.
For computer time, this research used GPU-based systems as
well as Shaheen supercomputer, both hosted at the Supercomputing Laboratory at King Abdullah University of Science and
Technology (KAUST).

\ifCLASSOPTIONcaptionsoff
  \newpage
\fi



\bibliographystyle{IEEEtran}
\bibliography{IEEEabrv,finalfile}
%



%

\begin{IEEEbiography}
	[{\includegraphics[width=1.25in,height=1.45in,clip,keepaspectratio]{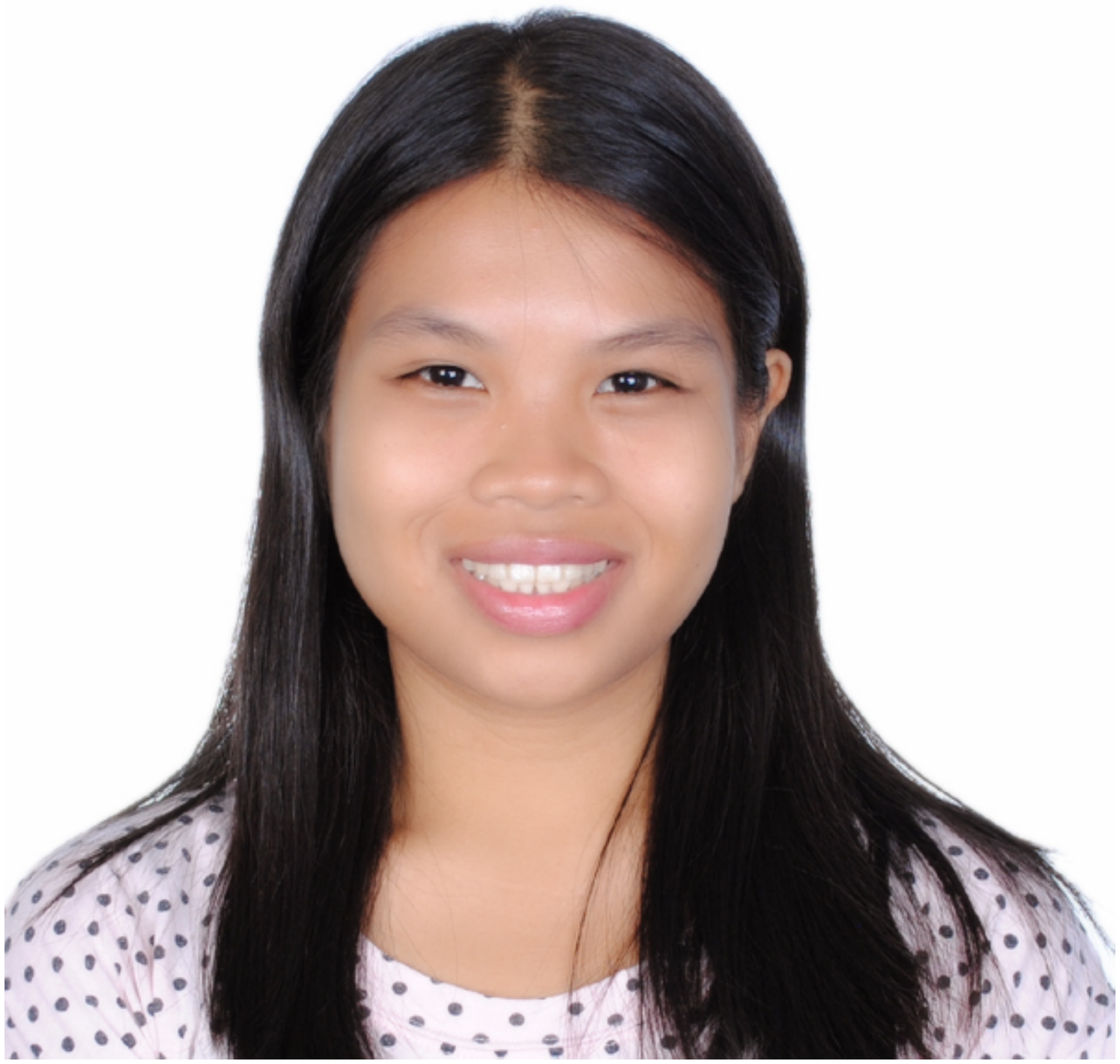}}]{Mary Lai Salva\~na}
	is a PhD student in the Spatio-Temporal Statistics \& Data Science group at King Abdullah University of Science and Technology (KAUST). She received her BS and MS degrees in Applied Mathematics in 2015 and 2016 from Ateneo de Manila University, Philippines. Her research interest includes multivariate spatio-temporal statistics and high-performance computing for large spatial and spatio-temporal datasets.
\end{IEEEbiography}
\vspace{-13mm}
\begin{IEEEbiography}
	[{\includegraphics[width=1in,height=1.25in,clip,keepaspectratio]{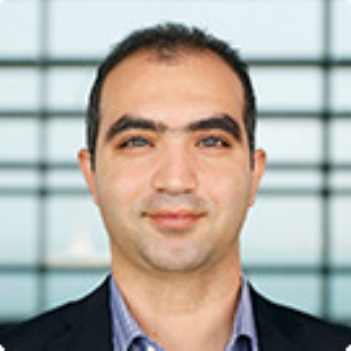}}]{Sameh Abdulah}
	is a research scientist at the Extreme Computing Research Center, 
	King Abdullah University of Science and Technology, Saudi Arabia. Sameh
	received his MS and PhD degrees from Ohio State University, 
	Columbus, USA, in 2014 and 2016, His work is centered around High
	Performance Computing (HPC) applications, bitmap indexing in big data, large spatial datasets, parallel statistical applications, 
	algorithm-based fault tolerance,  and Machine Learning and Data Mining algorithms.
\end{IEEEbiography}
\vspace{-10mm}
\begin{IEEEbiography}
	[{\includegraphics[width=1.25in,height=1.45in,clip,keepaspectratio]{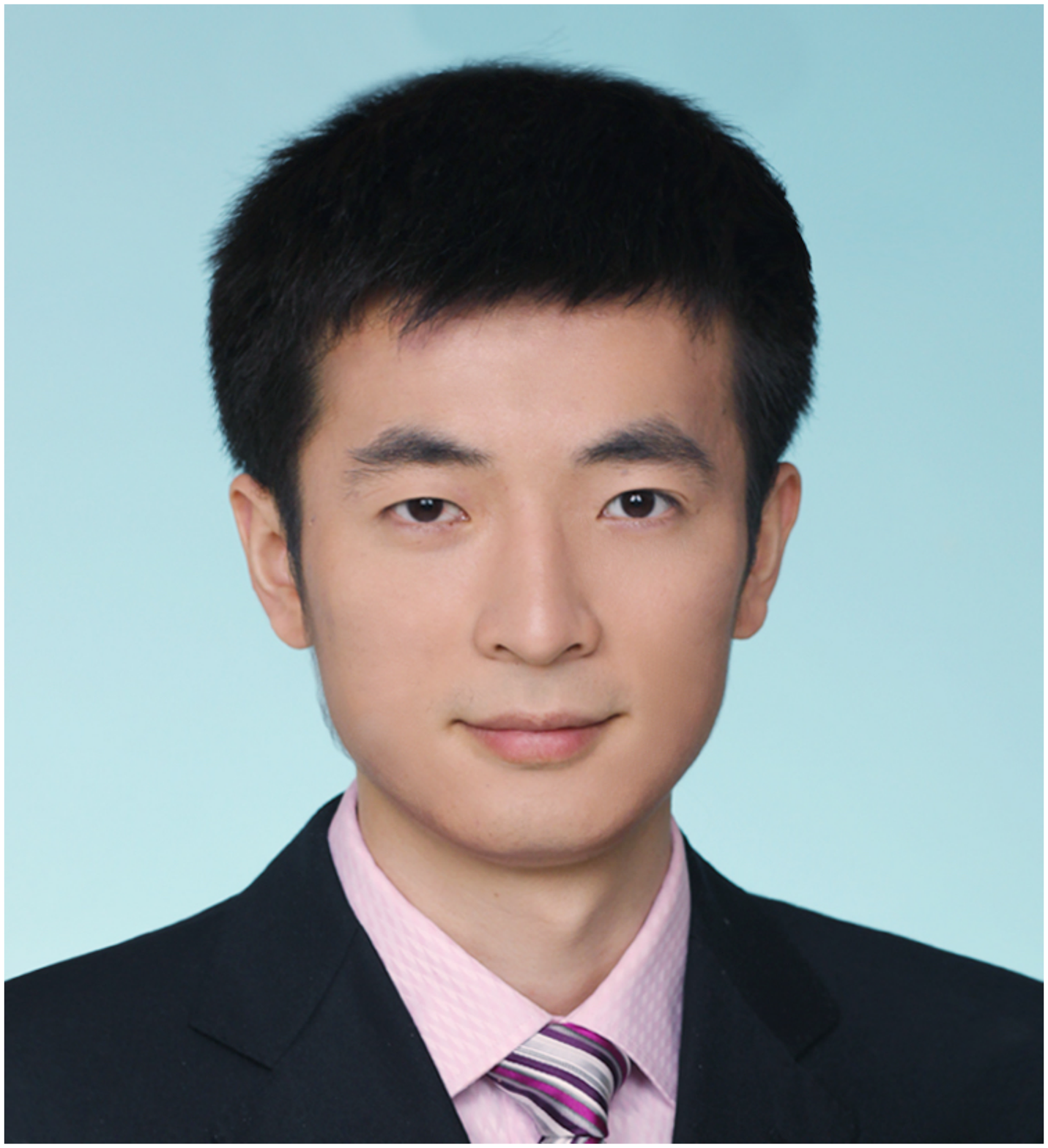}}]{Huang Huang}
	is a research scientist in the Spatio-Temporal Statistics \& Data Science group at King Abdullah University of Science and Technology (KAUST). Before working at KAUST, he did research on statistical computing for climate applications as a postdoc at the National Center for Atmospheric Research (NCAR), the Statistical and Applied Mathematical Sciences Institute (SAMSI), and Duke University. He received his Ph.D. in Statistics in 2017 from KAUST, master, and bachelor in Mathematics in 2014 and 2011 from Fudan University. His research interest includes spatio-temporal statistics, functional data analysis, Bayesian modeling, machine learning, and high-performance computing for large datasets.
\end{IEEEbiography}

\vspace{-10mm}

\begin{IEEEbiography}
	[{\includegraphics[width=1in,height=1.25in,clip,keepaspectratio]{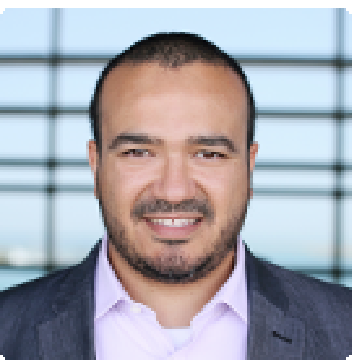}}]{Hatem Ltaief}
	is the Principal Research Scientist in the Extreme Computing
	Research Center at King Abdullah University of Science and Technology, Saudi Arabia.
	His research interests include parallel numerical
	algorithms, fault tolerant algorithms, parallel programming models,
	and performance optimizations for multicore architectures and hardware
	accelerators. His current research collaborators include Aramco,
	Total, Observatoire de Paris, NVIDIA, and Intel.
\end{IEEEbiography}

\vspace{-12mm}

\begin{IEEEbiography}
	[{\includegraphics[width=1in,height=1.25in,clip,keepaspectratio]{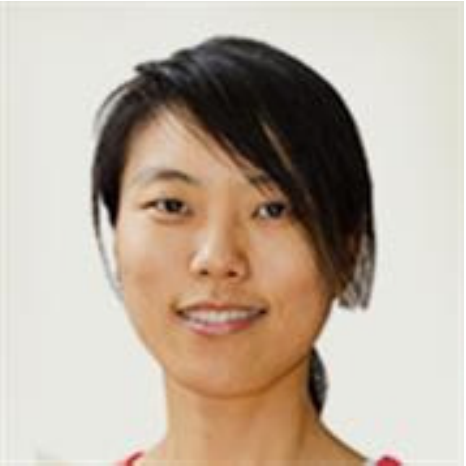}}]{Ying Sun}
received the PhD degree in statistics
from Texas A\&M University in 2011. She is an
associate professor of statistics with the King
Abdullah University of Science and Technology
(KAUST) in Saudi Arabia.
Her research interests include spatio-temporal statistics with
environmental applications, computational methods for large datasets,
uncertainty quantification and visualization, functional data analysis,
robust statistics, and statistics of extremes.
\end{IEEEbiography}

\vspace{-12mm}

\begin{IEEEbiography}
	[{\includegraphics[width=1in,height=1.25in,clip,keepaspectratio]{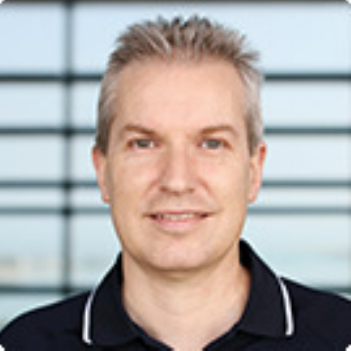}}]{Marc Genton}
received  the
PhD degree in statistics (1996) from the Swiss Federal Institute
 of Technology (EPFL), Lausanne. He is a distinguished professor of statistics
with the King Abdullah University of Science and
Technology (KAUST) in Saudi Arabia.
He is a
fellow of the American Statistical Association, of
the Institute of Mathematical Statistics, and the
American Association for the Advancement of
Science, and is an elected member of the International
Statistical Institute.
His research interests include statistical
analysis, flexible modeling, prediction, and uncertainty quantification of
spatio-temporal data, with applications in environmental and climate
science, renewable energies, geophysics, and marine science.
\end{IEEEbiography}
\vspace{-1cm}
\begin{IEEEbiography}
	[{\includegraphics[width=1in,height=1.25in,clip,keepaspectratio]{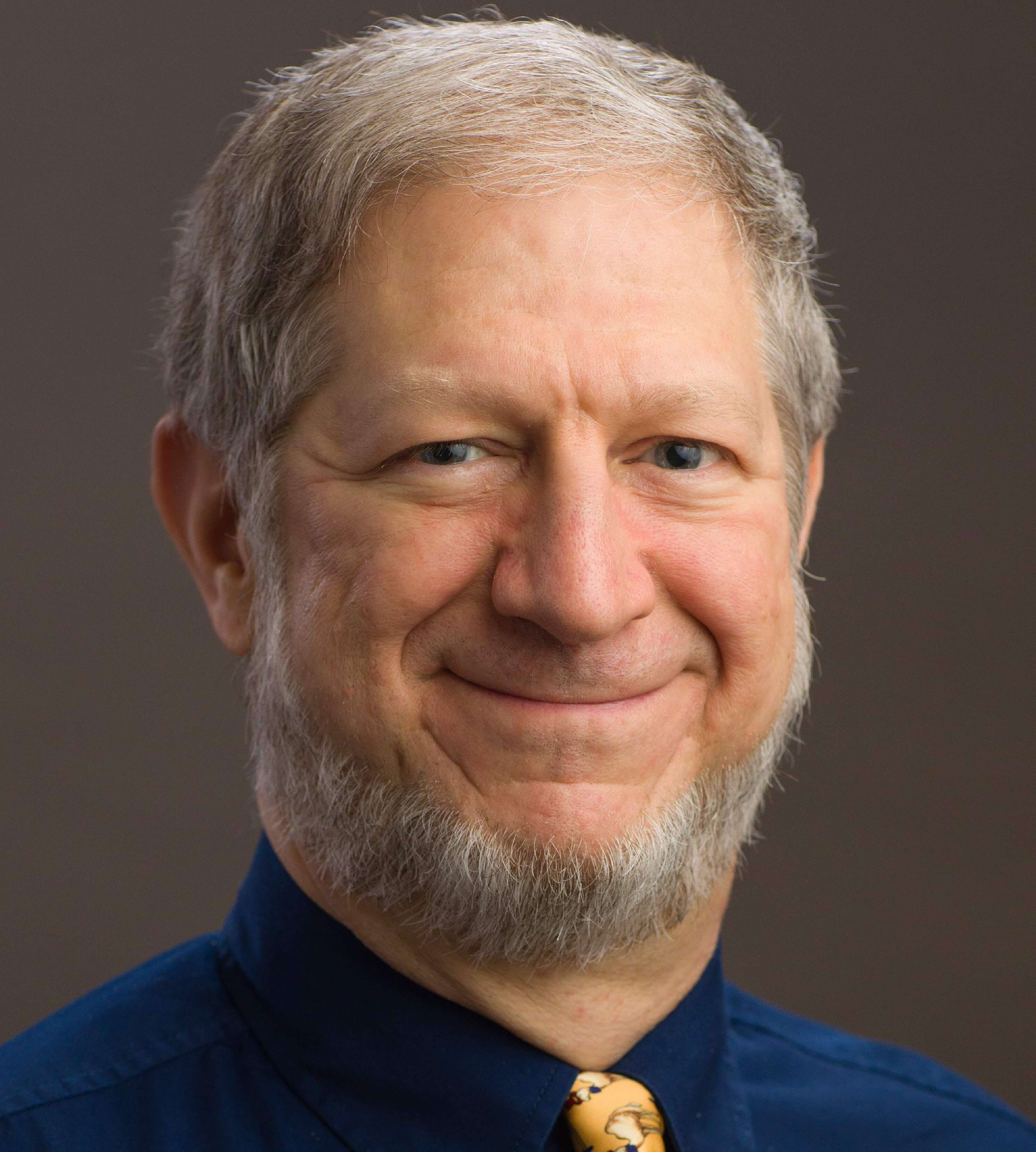}}]{David Keyes}
received the BSE degree
in aerospace and mechanical sciences from
Princeton University, in 1978, and the PhD
degree in applied mathematics from Harvard University,
in 1984.
He directs the Extreme Computing Research Center,
KAUST.
He works at the
interface between parallel computing and the
numerical analysis of PDEs, with a focus on
scalable implicit solvers. He helped develop and
popularize the Newton-Krylov-Schwarz (NKS),
Additive Schwarz Preconditioned Inexact Newton
(ASPIN), and Algebraic Fast Multipole (AFM) methods.
He is a fellow of the SIAM, AMS, and AAAS.

\end{IEEEbiography}




\end{document}